\documentclass[12pt]{article}

\usepackage{graphicx}
\usepackage{amsmath,amssymb}
\usepackage{bm}
\usepackage{graphicx,color}
\usepackage{wrapfig}
\begin{document}
\title{
\begin{flushright}
\ \\*[-80pt] 
\begin{minipage}{0.2\linewidth}
\normalsize
EPHOU-12-010 \\*[50pt]
\end{minipage}
\end{flushright}
{\Large \bf 
Minimal Neutrino Texture with Neutrino Mass Ratio and Cabibbo Angle
\\*[20pt]}}

\author{ 
\centerline{
Yusuke~Shimizu$^{1,}$\footnote{E-mail address: shimizu@muse.sc.niigata-u.ac.jp},
~Ryo~Takahashi$^{2,}$\footnote{E-mail address: takahashi@particle.sci.hokudai.ac.jp},~and
~Morimitsu~Tanimoto$^{1,}$\footnote{E-mail address: tanimoto@muse.sc.niigata-u.ac.jp} }
\\*[20pt]
\centerline{
\begin{minipage}{\linewidth}
\begin{center}
$^1${\it \normalsize
Department of Physics, Niigata University,~Niigata 950-2181, Japan } \\*[4pt]
$^2${\it \normalsize
Department of Physics, Faculty of Science, Hokkaido University, \\ 
Sapporo 060-0810, Japan } 
\end{center}
\end{minipage}}
\\*[70pt]}

\date{
\centerline{\small \bf Abstract}
\begin{minipage}{0.9\linewidth}
\medskip 
\medskip 
\small
We present neutrino mass matrix textures in a minimal framework of the type-I 
seesaw mechanism where two right-handed Majorana neutrinos are introduced 
in order to reproduce experimental results of neutrino oscillations. 
The textures can lead to experimentally favored leptonic mixing angles described 
by the tri-bimaximal mixing with one additional rotation. 
We present minimal and next to minimal textures for the normal mass hierarchy case 
in a context of the texture zero. A minimal texture in the inverted hierarchy case 
is also constructed, which does not have any vanishing entries in a Dirac neutrino 
mass matrix. We also discuss some cases that model parameters in the textures are 
supposed to be a neutrino mass ratio and/or the Cabibbo angle. 
Predicted regions of mixing angles, a leptonic CP-violation parameter, 
and an effective mass for the neutrino-less double beta decay are presented 
in all textures.
\end{minipage}
}

\begin{titlepage}
\maketitle
\thispagestyle{empty}
\end{titlepage}

\section{Introduction}

Neutrino oscillation experiments have revealed the lepton flavor to be large 
mixing, which are completely different from the quark mixing. 
Especially reactor experiments have observed a non-zero $\theta _{13}$, 
which is the last mixing angle of 
lepton sector~\cite{An:2012eh}.  
Now neutrino oscillation experiments go into a new phase 
of precise determinations of lepton mixing angles and neutrino mass squared
 differences~\cite{Tortola:2012te,Fogli:2012ua,GonzalezGarcia:2012sz}. 
Therefore, precise predictions are required for theoretical studies of
neutrino mixing angles and neutrino mass ratios.

Before the reactor experiments reported 
the non-zero value of $\theta _{13}$, we had a 
paradigm of "tri-bimaximal mixing" (TBM)~\cite{Harrison:2002er,Harrison:2002kp}, 
which is a simple mixing pattern for leptons and can be easily derived from flavor symmetries. 
Actually many authors have discussed the TBM by introducing
non-Abelian discrete flavor symmetries~\cite{Ishimori:2010au,Ishimori:2012zz}.  
However, the non-vanishing $\theta _{13}$ forces to study a deviation from 
the TBM~\cite{Rodejohann:2012cf,Damanik:2012uw} or other patterns of lepton mixing angles, 
e.g. tri-bimaximal-Cabibbo mixing~\cite{King:2012vj}.\footnote{
See also \cite{Shimizu:2010pg} for a discussion of deviation from the TBM and a 
quark-lepton complementarity in a model independent way.}

Now it is necessary to obtain simple textures of a neutrino mass matrix 
in order to investigate the origin
 of lepton mixing angles with the non-zero $\theta _{13}$. 
Since the exact TBM cannot explain the current results of neutrino oscillation experiments, 
we consider two patterns of deviation from the TBM, 
which are defined by 
an additional rotation of 1-3 or 2-3 generation to the TBM.\footnote{A rotation of
1-2 generation still leads to $\theta _{13}=0$, 
which has been just ruled out by the recent neutrino oscillation results.} 
These two patterns give a different prediction for 
the magnitude of $\sin ^2\theta _{12}$. 
The additional 1-3 rotation strictly leads to $\sin ^2\theta _{12}>1/3$. 
On the other hand, in the case of 2-3 rotation, 
the additional rotation can give rise to $\sin ^2\theta _{12}<1/3$, 
which is in favor for experimental results. 
Therefore, we will focus on the additional 2-3 rotation to the TBM. 

In this paper, we show how to derive the desired neutrino mixing pattern 
in a context of the type-I seesaw mechanism~\cite{seesaw}, 
which can give a natural realization of tiny neutrino masses 
compared to masses of other standard model (SM) fermions, with three generations
 of right-handed Majorana neutrinos in addition to three flavors of left-handed 
Majorana neutrinos in the SM. The presence of three generations of right-handed 
Majorana neutrinos are required for a gauge anomaly cancellation when one introduces a gauged 
$U(1)_{B-L}$ symmetry as a physics beyond the SM, e.g. SO(10) grand unified 
theory where the right-handed Majorana neutrinos can be naturally embedded. Models with 
three right-handed Majorana neutrinos are also well motivated for simultaneous 
explanations of important unsolved mysteries in the current particle physics and
 cosmology. They are, for example, a generation of baryon asymmetry of the 
Universe (BAU), to give a dark matter (DM) candidate, and an explanation of 
LSND/MiniBOONE anomaly in addition to a realization of the tiny neutrino masses 
(e.g. see~\cite{Asaka:2005an}-\cite{Chen:2011ai}). 
It is known that in scenarios including a keV sterile neutrino DM, 
one of three right-handed Majorana neutrinos (which is a DM candidate with the mass of keV) 
should not affect the active neutrino masses in order to satisfy cosmological bounds. 
Therefore, we can effectively write down the right-handed Majorana neutrino mass matrix 
$M_R$ and the Dirac neutrino mass matrix $M_D$ in terms of sub-matrices 
in some scenarios including three generations of the right-handed Majorana neutrinos 
(one of them is a DM candidate) as follows: 
\begin{equation}
M_R^{3\times 3}=
\begin{pmatrix}
M_{R}^{1\times 1} & 0 \\
0 & M_{R}^{2\times 2}
\end{pmatrix},\quad M_D=\left (Y^{3\times 1}_D\quad Y^{3\times 2}_D\right )v,
\label{1-2}
\end{equation}
where $M_R^{1\times 1}$ is the right-handed Majorana neutrino mass, 
which is the keV scale in the sterile neutrino DM scenario, 
$M_{R}^{2\times 2}$ is a $2\times 2$ right-handed Majorana neutrino mass matrix, $Y^{3\times 1}_D$ 
is a $3\times 1$ Dirac neutrino Yukawa matrix among the SM and the right-handed Majorana neutrino 
with mass of $M_{R}^{1\times 1}$, $Y^{3\times 2}_D$ is a $3\times 2$ Dirac neutrino Yukawa one, 
and $v$ is a vacuum expectation value (VEV) of the SM Higgs, respectively. 
Here, note that structures of mass spectrum and Yukawa matrix 
given in the Eq.~(\ref{1-2}) are naively splitting. Such kind of split mass spectrum
 and Yukawa matrix can be naturally realized in a split seesaw 
mechanism~\cite{Kusenko:2010ik}, 
Froggatt-Nielsen mechanism~\cite{Froggatt:1978nt,Merle:2011yv}, or a flavor 
symmetry~\cite{Lindner:2010wr}. Then, by using the seesaw mechanism, 
the left-handed Majorana neutrino mass matrix can be separately given as, 
\begin{equation}
M_\nu =\left[Y^{3\times 2}_D\left (M_{R}^{2\times 2}\right )^{-1}\left (Y_D^{3\times 2}\right )^T
+Y_D^{3\times 1}\left (M_R^{1\times 1}\right )^{-1}(Y_D^{3\times 1})^T\right ]v^2.
\label{split-seesaw}
\end{equation}

It is well known that an introduction of two right-handed Majorana neutrinos is a minimal scheme 
in order to reproduce the experimental data of three leptonic flavor mixing angles 
and two mass squared differences of the 
neutrinos~\cite{King:1998jw}-\cite{Bhattacharya:2006aw}. Such situation can be 
embedded into the above scenario described by the Eqs.~(\ref{1-2}) and 
(\ref{split-seesaw}) without the loss of generality. For instance, 
if one consider the lightest sterile neutrino $M_R^{1\times 1}$ as a candidate of the 
DM with the keV mass, the second term of the Eq.~(\ref{split-seesaw}) does not 
affect the flavor mixing and the neutrino masses.\footnote{There is another 
option, i.e. one of the sterile neutrinos are super-heavy compared with the 
other two ones. Also in the case, such a super-heavy sterile neutrino does not 
contribute to the flavor mixing angles and two mass scales of the neutrinos 
since the sterile neutrino is decoupled from the theory at a high energy.} 
Therefore, only the first term of the Eq.~(\ref{split-seesaw}) contributes to 
the mixing and the masses. In other words, the neutrino flavor mixing is 
determined by only the structures of the $3\times 2$ Dirac neutrino mass matrix
 and the $2\times 2$ right-handed Majorana neutrino one. As the consequences, 
the lightest left-handed Majorana neutrino mass is very tiny compared with the other 
ones. In these setup, we will investigate the neutrino mass matrix texture 
which can reproduce the current neutrino oscillation experiments containing the
 result of the non-zero $\theta _{13}$.\footnote{See also~\cite{Adulpravitchai:2011rq} 
for discussions of the flavor mixing in the context of the split seesaw mechanism.}

The paper is organized as follows. We discuss some patterns of the deviation 
from the TBM in section~2. In section~3, we construct the minimal texture leading 
to the additional 2-3 rotation to the TBM. Finally, we present some specific 
cases of the minimal texture where model parameters are taken as the ratio 
between two mass squared differences of the neutrinos and/or the Cabibbo angle 
in section 4. Section 5 is devoted to the summary. 

\section{Deviations from the tri-bimaximal mixing}

Mixing matrix for quarks and leptons is independently given by different unitary matrices including 
three mixing angles $\theta _{ij}$ $(i,j=1,2,3;~i<j)$ and one Dirac phase $\delta $. 
One of well-known parameterization of the mixing matrix is the PDG one~\cite{Beringer:1900zz} 
described as
\begin{align}
U &\equiv 
\begin{pmatrix}
1 & 0 & 0 \\
0 & c_{23} & s_{23} \\
0 & -s_{23} & c_{23}
\end{pmatrix} 
\begin{pmatrix}
c_{13} & 0 & s_{13}e^{-i\delta } \\
0 & 1 & 0 \\
-s_{13}e^{i\delta } & 0 & c_{13}
\end{pmatrix}
\begin{pmatrix}
c_{12} & s_{12} & 0 \\
-s_{12} & c_{12} & 0 \\
0 & 0 & 1
\end{pmatrix} \nonumber \\
&= 
\begin{pmatrix}
c_{12} c_{13} & s_{12} c_{13} & s_{13}e^{-i\delta } \\
-s_{12} c_{23} - c_{12} s_{23} s_{13}e^{i\delta } & 
c_{12} c_{23} - s_{12} s_{23} s_{13}e^{i\delta } & s_{23} c_{13} \\
s_{12} s_{23} - c_{12} c_{23} s_{13}e^{i\delta } & 
-c_{12} s_{23} - s_{12} c_{23} s_{13}e^{i\delta } & c_{23} c_{13}
\end{pmatrix}, \label{U}
\end{align}
where $c_{ij}$ and  $s_{ij}$ denote $\cos \theta _{ij}$ and $\sin \theta _{ij}$, respectively. 
Throughout this work we focus only on the lepton mixing matrix, namely 
Pontecorvo-Maki-Nakagawa-Sakata (PMNS) matrix $U_{\text{PMNS}}$~\cite{Maki:1962mu,Pontecorvo:1967fh}. 
If neutrinos are Majorana particles, Majorana phases are included in the 
 left-handed Majorana neutrino masses. 
For the leptonic mixing matrix, 
Harrison-Perkins-Scott proposed a simple form of the mixing pattern, so-called the 
tri-bimaximal mixing (TBM)~\cite{Harrison:2002er,Harrison:2002kp}, as 
\begin{equation}
V_{\text{TBM}}=
\begin{pmatrix}
\frac{2}{\sqrt{6}} & \frac{1}{\sqrt{3}} & 0 \\
-\frac{1}{\sqrt{6}} & \frac{1}{\sqrt{3}} & -\frac{1}{\sqrt{2}} \\
-\frac{1}{\sqrt{6}} & \frac{1}{\sqrt{3}} & \frac{1}{\sqrt{2}}
\end{pmatrix},
\label{TBM}
\end{equation}
before reporting the non-vanishing $\theta _{13}$. This TBM matrix reads 
\begin{equation}
|U_{e2}|=\frac{1}{\sqrt{3}},\qquad |U_{e3}|=0,\qquad 
|U_{\mu 3}|=\frac{1}{\sqrt{2}}.
\end{equation}
Since the TBM pattern is suggestive for studies of flavor physics behind 
the SM, a large number of works have been proposed for the realization of the 
TBM. However, the TBM has been just excluded due to the recent 
experimental result of the non-vanishing $\theta _{13}$~\cite{An:2012eh}. 
Therefore, one should consider alternatives of the TBM to reproduce the
 experimental results. One of the simple direction to explain the experimental data 
might be to minimally extend the TBM. The first task in this direction 
is to realize the non-zero $\theta _{13}$ because the experimentally observed mixing 
angles of $\theta _{12}$ and $\theta _{23}$ are still well approximated by the 
TBM. 

In order to get the non-zero $\theta _{13}$, one can consider an additional rotation of 
$1$--$3$ generation. 
With this additional rotation, the PMNS mixing matrix can be written as 
\begin{equation}
U_{\text{PMNS}}=V_{\text{TBM}}
\begin{pmatrix}
\cos \phi & 0 & \sin \phi \\
0 & 1 & 0 \\
-\sin \phi & 0 & \cos \phi 
\end{pmatrix}.
\end{equation}
Then, magnitudes of relevant mixing matrix elements 
to each mixing angle are 
\begin{equation}
\left |U_{e2}\right |=\frac{1}{\sqrt{3}},\quad 
\left |U_{e3}\right |=\left |\frac{2\sin \phi }{\sqrt{6}}\right |,\quad
\left |U_{\mu 3}\right |=\left |-\frac{\sin \phi }{\sqrt{6}}+\frac{\cos \phi }{\sqrt{2}}
\right |~.
\end{equation}
It can be seen that $\sin ^2\theta _{12}>1/3$ is predicted 
in this case,\footnote{
This case can be easily realized in some actual flavor 
models~\cite{Ishimori:2010fs,Shimizu:2011xg}.
} while the best fit value of $\sin ^2\theta _{12}$ in a global analysis 
of the neutrino oscillation experiments tends to be lower than $1/3$~\cite{Tortola:2012te}.

Another possibility of deviation from the TBM by an additional rotation is the $2$--$3$ rotation 
from the TBM~\cite{Rodejohann:2012cf}. 
In the case, the PMNS mixing matrix is written as 
\begin{equation}
U_{\text{PMNS}}=V_{\text{TBM}}
\begin{pmatrix}
1 & 0 & 0 \\
0 & \cos \phi & \sin \phi \\
0 & -\sin \phi & \cos \phi 
\end{pmatrix},
\label{23rotation}
\end{equation}
and the relevant mixing matrix elements are
\begin{equation}
\left |U_{e2}\right |=\left |\frac{\cos \phi }{\sqrt{3}}\right |,\quad 
\left |U_{e3}\right |=\left |\frac{\sin \phi }{\sqrt{3}}\right |,\quad
\left |U_{\mu 3}\right |=
\left |\frac{\sin \phi }{\sqrt{3}}+\frac{\cos \phi }{\sqrt{2}}\right |~.
\end{equation}
We find the upper limit on the magnitude of $\theta _{12}$ as 
$\sin ^2\theta _{12}<1/3$, which is in favor of 
the current experimental results. 

There also exists the last possibility of deviation constructed by an additional
 rotation to the TBM, i.e. 1-2 rotation, but this case does not lead to the non-zero
 $\theta _{13}$. Therefore, we will focus on the additional 2-3 rotation and 
construct neutrino mass textures leading to the rotation, in the ($2+1$) framework
of the right-handed Majorana neutrinos described by the Eqs.~(\ref{1-2}) and 
(\ref{split-seesaw}) in the following sections. 


\section{Towards the minimal texture}
We investigate neutrino mass matrix textures, which can lead to the additional 2-3 
rotation to the TBM for the PMNS matrix, in the ($2+1$) framework 
described by the Eqs.~(\ref{1-2}) and (\ref{split-seesaw}). Such framework can 
be applied to general models with the keV sterile neutrino DM, and required 
hierarchical mass spectrum of the right-handed Majorana neutrinos and the Yukawa 
couplings are partially realized by some mechanisms, e.g. the split seesaw 
mechanism~\cite{Kusenko:2010ik}. As we mentioned above, the $2\times 2$ 
right-handed Majorana neutrino mass matrix and the $3\times 2$ Dirac neutrino mass 
matrix are effectively enough to give the flavor mixing and two mass squared 
differences of the neutrinos. Therefore, we focus on the structures of 
$M_{R}^{2\times 2}$ and $Y^{3\times 2}_D$ in the $(2+1)$ framework, the 
Eqs.~(\ref{1-2}) and (\ref{split-seesaw}). 

We can take the $2\times 2$ right-handed Majorana neutrino mass matrix $M_R^{2\times 2}$ 
to be diagonal in general as follow: 
\begin{equation}
M_R^{2\times2}=m_R
\begin{pmatrix}
\frac{1}{p} & 0 \\
0 & 1
\end{pmatrix},
\end{equation}
where, $m_R$ is a fundamental mass scale of the 
right-handed Majorana neutrino and $p$ is a 
ratio between the two right-handed Majorana neutrino masses. 
On the other hand, the relevant Dirac neutrino mass matrix 
$M_D^{3\times 2}$ is defined as 
\begin{equation}
M_D^{3\times 2}\equiv Y_D^{3\times 2}v=
\begin{pmatrix}
a & d \\
b & e \\
c & f
\end{pmatrix}v,
\end{equation}
where $a\sim f$ are Yukawa couplings. 
By using the seesaw mechanism, the left-handed Majorana neutrino mass matrix $M_{\nu }$ 
can be well approximated by 
\begin{equation}
M_{\nu }=M_DM_R^{-1}M_D^T
\simeq M_D^{3\times2}(M_R^{2\times2})^{-1}(M_D^{3\times2})^T=
\frac{v^2}{m_R} 
\begin{pmatrix}
a^2p+d^2 & abp+de & acp+df \\
abp+de & b^2p+e^2 & bcp+ef \\
acp+df & bcp+ef & c^2p+f^2
\end{pmatrix},
\label{left-handed-Majorana-1}
\end{equation}
in the decoupling limit of $M_R^{1\times 1}$ for the light neutrino mass matrix 
Eq.~(\ref{split-seesaw}).
By performing the TBM matrix to the neutrino mass matrix $M_\nu $, the 
left-handed Majorana neutrino mass matrix is rewritten as 
\begin{align}
&M_\nu =\frac{v^2}{m_R}V_{\text{TBM}}^T
\begin{pmatrix}
\frac{A^2p+D^2}{6} & \frac{ABp+DE}{3\sqrt{2}} & \frac{ACp+DF}{2\sqrt{3}} \\
\frac{ABp+DE}{3\sqrt{2}} & \frac{B^2p+E^2}{3} & \frac{BCp+EF}{\sqrt{6}} \\
\frac{ACp+DF}{2\sqrt{3}} & \frac{BCp+EF}{\sqrt{6}} & \frac{C^2p+F^2}{2}
\end{pmatrix}V_{\text{TBM}},
\label{left-handed-Majorana-TBM}
\end{align}
where 
\begin{align}
A\equiv 2a-b-c,~~B\equiv a+b+c,~~C\equiv c-b,~~D\equiv 2d-e-f,~~E\equiv d+e+f,~~F\equiv f-e.
\end{align}
The Eq.~(\ref{left-handed-Majorana-TBM}) is called as the matrix in the TBM 
basis. We discuss neutrino mass structures to realize the additional 2-3 
rotation for both cases of the normal and inverted neutrino mass hierarchies in
 sections 3.1 and 3.2, respectively. 

\subsection{Normal neutrino mass hierarchy}
At first, we consider the case of the normal neutrino mass hierarchy (NH). 
It is seen that conditions 
for the additional 2-3 rotation are 
\begin{equation}
A=2a-b-c=0,\qquad D=2d-e-f=0.
\end{equation}
After imposing these conditions on the Eq.~(\ref{left-handed-Majorana-TBM}), 
the mass matrix is rewritten as 
\begin{equation}
M_\nu =\frac{v^2}{m_R}V_{\text{TBM}}^T
\begin{pmatrix}
0 & 0 & 0 \\
0 & \frac{3}{4}\left ((b+c)^2p+(e+f)^2\right ) & 
\frac{1}{2}\sqrt{\frac{3}{2}}\left ((c^2-b^2)p-e^2+f^2\right ) \\
0 & \frac{1}{2}\sqrt{\frac{3}{2}}\left ((c^2-b^2)p-e^2+f^2\right ) & 
\frac{1}{2}\left ((b-c)^2p+(e-f)^2\right )
\end{pmatrix}V_{\text{TBM}}.
\label{left-handed-Majorana-general}
\end{equation}
Since the Majorana neutrino mass can be rescaled in the seesaw formula,
 the right-handed Majorana and Dirac neutrino mass matrices are written by 
putting $p=1$ as 
\begin{equation}
M_R^{2\times 2}=m_R
\begin{pmatrix}
1 & 0 \\
0 & 1
\end{pmatrix},\qquad 
M_D^{3\times 2}=
\begin{pmatrix}
\frac{b+c}{2} & \frac{e+f}{2} \\
b & e \\
c & f
\end{pmatrix}v,
\label{general-texture}
\end{equation}
respectively.\footnote{
If one discuss a phenomenology depending on the right-handed Majorana neutrino masses, 
e.g. a generation mechanism of the BAU, one cannot rescale. However, such rescaling 
does not change our results of analyses for the mixing angles.} Starting from 
these textures, we possibly simplify them in a context of texture zeros, which 
is one of attractive strategies to discuss minimal texture. 

\subsubsection{Two zero texture} 
Three or more texture zeros in the $3\times2$ Dirac neutrino mass matrix described by 
the Eq.~(\ref{general-texture}) never give realistic lepton mixing angles consistent 
with the experimentally observed ones. Therefore, we begin with two zero texture. 
There are three cases for the two zero texture, in which possible conditions are 
\begin{equation}
\text{(i)}~b+c=0,~f=0,\quad \text{(i\hspace{-.1em}i)}~b+c=0,~e=0,\quad 
\text{(i\hspace{-.1em}i\hspace{-.1em}i)}~c=0,~e=0.
\end{equation}
The other conditions give one vanishing mixing angle and lead only one 
non-vanishing neutrino mass. Now, corresponding Dirac neutrino mass matrices 
are given as 
\begin{equation}
\frac{M_D^{3\times 2}}{v}=\left\{
\begin{array}{cl}
\begin{pmatrix}
0 & \frac{e}{2} \\
b & e \\
-b & 0
\end{pmatrix} & \mbox{ for (i) $b+c=0,~f=0$}\\ 
\begin{pmatrix}
0 & \frac{f}{2} \\
b & 0 \\
-b & f
\end{pmatrix} & \mbox{ for (i\hspace{-.1em}i) $b+c=0,~e=0$}\\
\begin{pmatrix}
\frac{b}{2} & \frac{f}{2} \\
b & 0 \\
0 & f
\end{pmatrix} & \mbox{ for (i\hspace{-.1em}i\hspace{-.1em}i) $c=0,~e=0$}
\end{array}
\right ..
\label{MD}
\end{equation}
It is also seen that resultant mixing angles will not be changed under the 
rescaling of $p=1$ in the right-handed Majorana neutrino mass matrix and replacement 
between the first and the second columns of the Yukawa matrices in the 
right-hand side of the Eq.~(\ref{MD}), but such replacement will lead to a 
wrong mass ratio $m_2/m_3>1$ as we will see later. Therefore, all patterns are 
included in the above three cases to be consistent with the experimental 
results.

We can express the $3\times 2$ Dirac neutrino mass matrix in terms of 
only one parameter $e$ or $f$ 
by rescaling the overall factor $v$ to $m_0$ and redefining rescaled 
$e\rightarrow e'=em_0/v$ or $f\rightarrow f'=fm_0/v$ as $e(\equiv e')$ and 
$f(\equiv f')$ without the loss of generality, 
\begin{equation}
\frac{M_D^{3\times2}}{m_0}=
\left \{
\begin{array}{cl}
\begin{pmatrix}
0 & \frac{e}{2} \\
\frac{1}{\sqrt{2}} & e \\
-\frac{1}{\sqrt{2}} & 0
\end{pmatrix} & \mbox{ for (i) $b+c=0,~f=0$}\\
\begin{pmatrix}
0 & \frac{f}{2} \\
\frac{1}{\sqrt{2}} & 0 \\
-\frac{1}{\sqrt{2}} & f
\end{pmatrix} & \mbox{ for (i\hspace{-.1em}i) $b+c=0,~e=0$}\\
\begin{pmatrix}
\frac{1}{2} & \frac{f}{2} \\
1 & 0 \\
0 & f
\end{pmatrix} & \mbox{ for (i\hspace{-.1em}i\hspace{-.1em}i) $c=0,~e=0$}
\end{array}
\right ..
\end{equation}

\newpage 
\noindent\underline{{\bf (i) {\boldmath $b+c=0$ and $f=0$} case}}\\

\begin{figure}
\begin{center}
\begin{tabular}{cc}
\hline 
\multicolumn{2}{c}{Two zero texture in the NH} \\
\hline \hline 
\multicolumn{2}{c}{(i) $b+c=0$ and $f=0$ case} \\
\hline \\
\hspace{1.1cm}(a) & \hspace{1cm}(b) \\
\includegraphics[width=6.6cm]{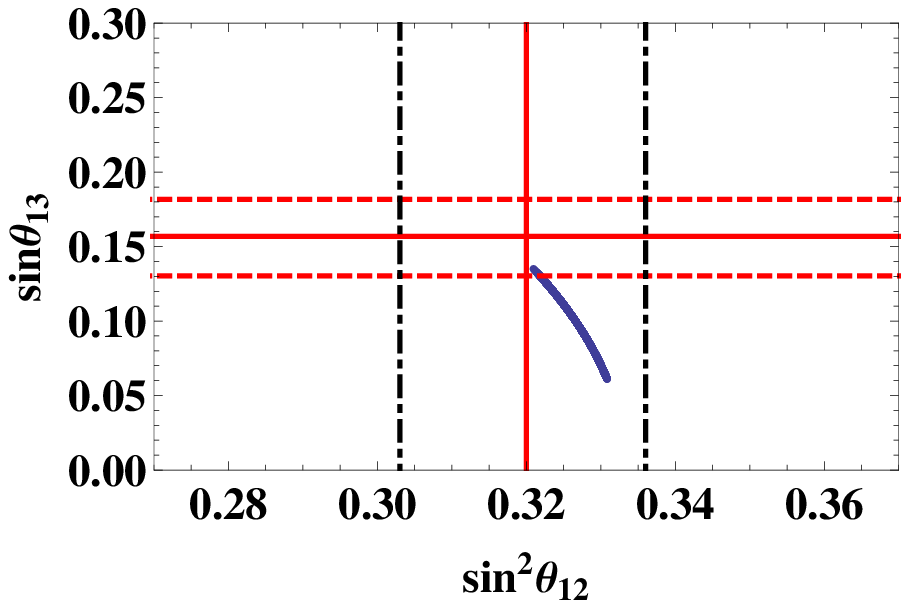} & \includegraphics[width=6.6cm]{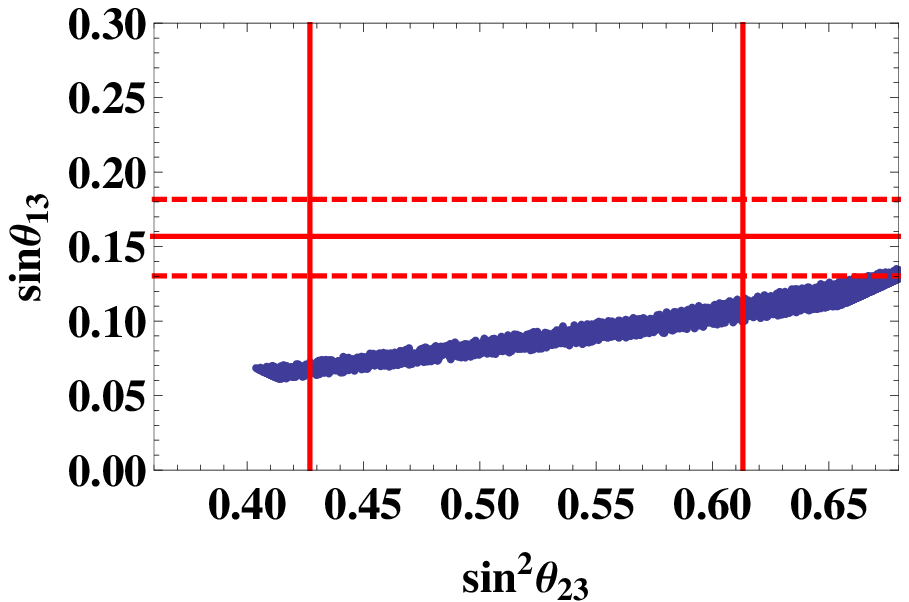} \\
\\
\hspace{1.1cm}(c) & \hspace{1cm}(d) \\
\includegraphics[width=6.6cm]{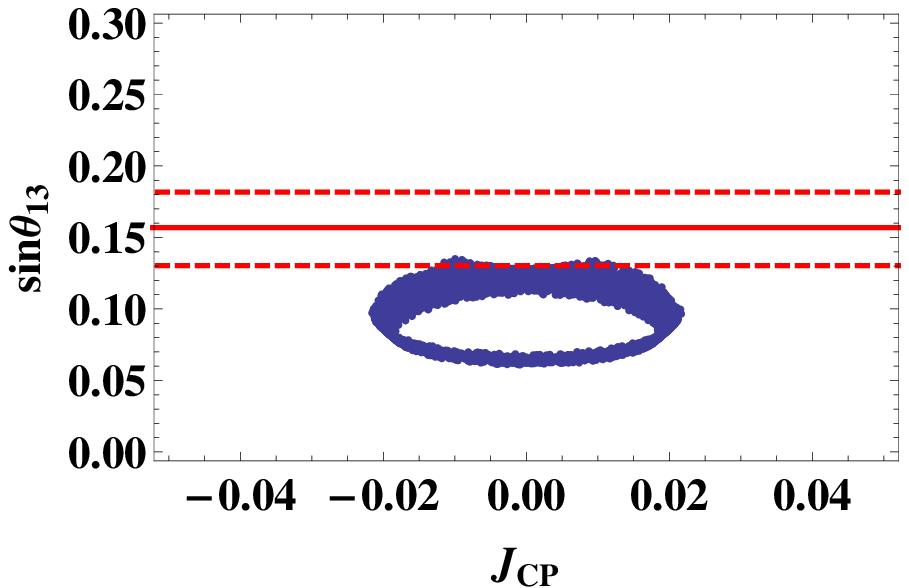} & \includegraphics[width=6.6cm]{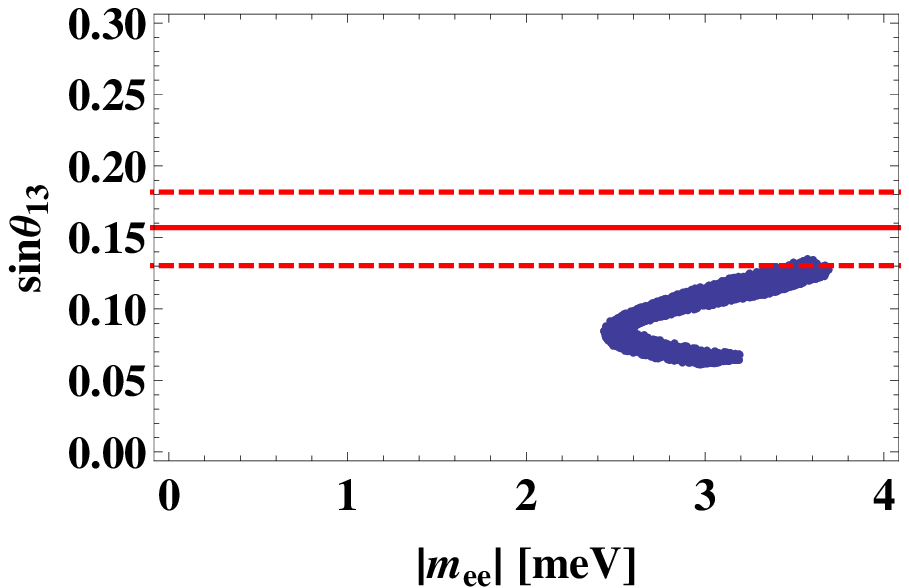} \\
\\
\multicolumn{2}{c}{\hspace{1cm}(e)}\\
\multicolumn{2}{c}{\includegraphics[width=6.6cm]{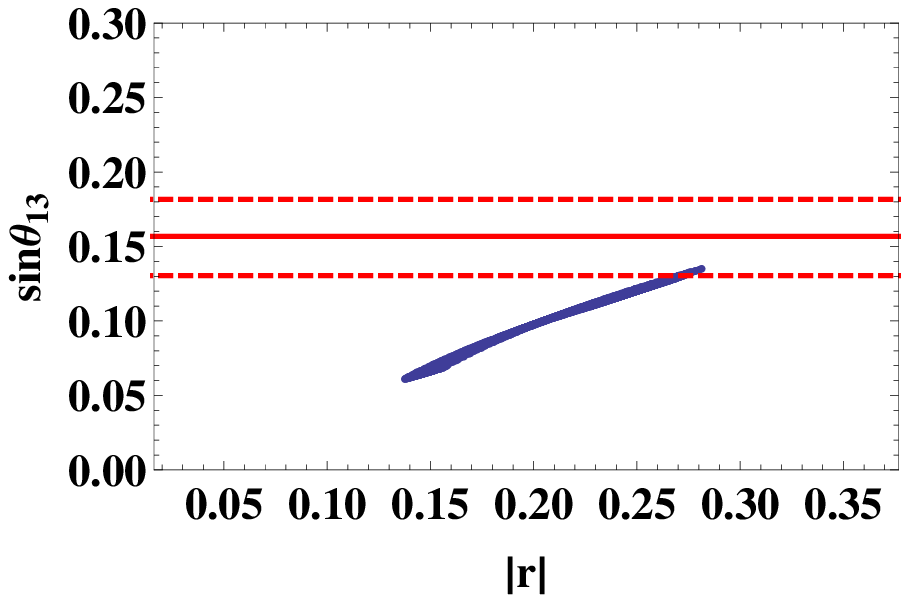}} \\
\\
\hline
\end{tabular}
\end{center}
\caption{Predicted regions of mixing angles, the Jarlskog invariant, the 
effective neutrino mass for the $0\nu \beta \beta $, and a favored region of $|r|$ 
in the case (i) of NH with two zero texture: (a) 
$\sin ^2\theta _{12}$--$\sin \theta _{13}$, (b) 
$\sin ^2\theta _{23}$--$\sin \theta _{13}$, (c) $J_{CP}$--$\sin \theta _{13}$, (d) 
$|m_{ee}|$--$\sin \theta _{13}$, and (e) $|r|$--$\sin \theta _{13}$ planes. 
In the figures (a) and (b), the plots are within $3\sigma $ of the 
$\sin ^2\theta _{23}$ and $\sin ^2\theta _{12}$, respectively. In the figures (c), 
(d), and (e), the plots are within $3\sigma$ ranges of both 
$\sin ^2\theta _{23}$ and $\sin ^2\theta _{12}$. The best fit values of 
experimental data are denoted by solid lines in all figures. The chained and 
dashed lines denote experimental bounds of $1\sigma $ and $3\sigma $, 
respectively.}
\label{fig1}
\end{figure}

In the case (i), the left-handed Majorana neutrino is obtained as 
\begin{equation}
M_\nu =\frac{m_0^2}{m_R}V_{\text{TBM}}^T
\begin{pmatrix}
0 & 0 & 0 \\
0 & \frac{3}{4}e^2 & -\frac{1}{2}\sqrt{\frac{3}{2}}e^2 \\
0 & -\frac{1}{2}\sqrt{\frac{3}{2}}e^2 & 1+\frac{1}{2}e^2
\end{pmatrix}V_{\text{TBM}}.
\end{equation}
It can be seen that the additional 2-3 mixing angle, $\phi $ in the Eq.~(\ref{23rotation}), 
is too large to be consistent with the experimental data if $e\sim \mathcal{O}(1)$. 
Therefore, we should take as $e\ll 1$. Then, the neutrino mass eigenvalues are 
\begin{equation}
m_1=0,\quad \frac{m_2}{m_3}\simeq \frac{3}{4}e^2\equiv r,
\label{mass-2texture1}
\end{equation}
at the leading order, where $r$ is around a ratio between 
two left-handed Majorana neutrino mass eigenvalues. 
The additional 2-3 mixing angle is 
\begin{equation}
\tan (2\phi )\simeq -\sqrt{\frac{3}{2}}e^2,
\label{mixing-2texture1}
\end{equation}
where $e$ is taken to be real for simplicity. The relevant mixing matrix 
elements are written from the Eqs.~(\ref{mass-2texture1}) and 
(\ref{mixing-2texture1}) as 
\begin{equation}
\left |U_{e2}\right |\simeq \frac{1}{\sqrt{3}}\sqrt{1-\frac{2}{3}r^2},\quad 
\left |U_{e3}\right |\simeq \left |-\frac{\sqrt{2}}{3}r\right |,\quad 
\left |U_{\mu 3}\right |\simeq 
\left |-\frac{\sqrt{2}}{3}r+\frac{1}{\sqrt{2}}\sqrt{1-\frac{2}{3}r^2}\right |~.
\end{equation}
We can find that the three mixing angles are correlated through the parameter $r$. 
The correlations among three mixing angles and predicted regions 
for each mixing angle are numerically shown in Fig.~\ref{fig1}. 
\begin{figure}
\begin{center}
\begin{tabular}{cc}
\hline
\multicolumn{2}{c}{Two zero texture in the NH} \\
\hline \hline 
\multicolumn{2}{c}{(i\hspace{-.1em}i) $b+c=0$ and $e=0$ case} \\
\hline \\
\hspace{1.1cm}(a) & \hspace{1cm}(b) \\
\includegraphics[width=6.8cm]{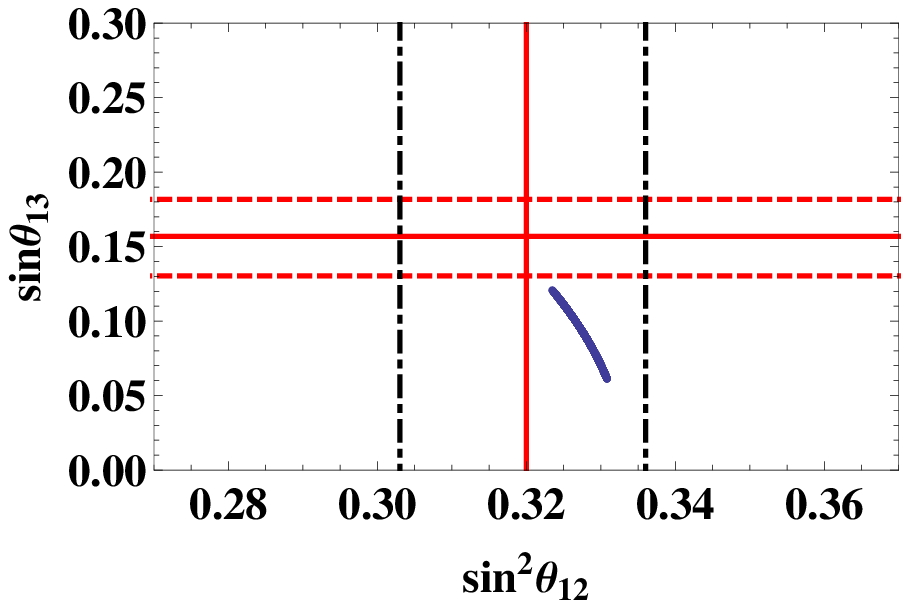} & \includegraphics[width=6.8cm]{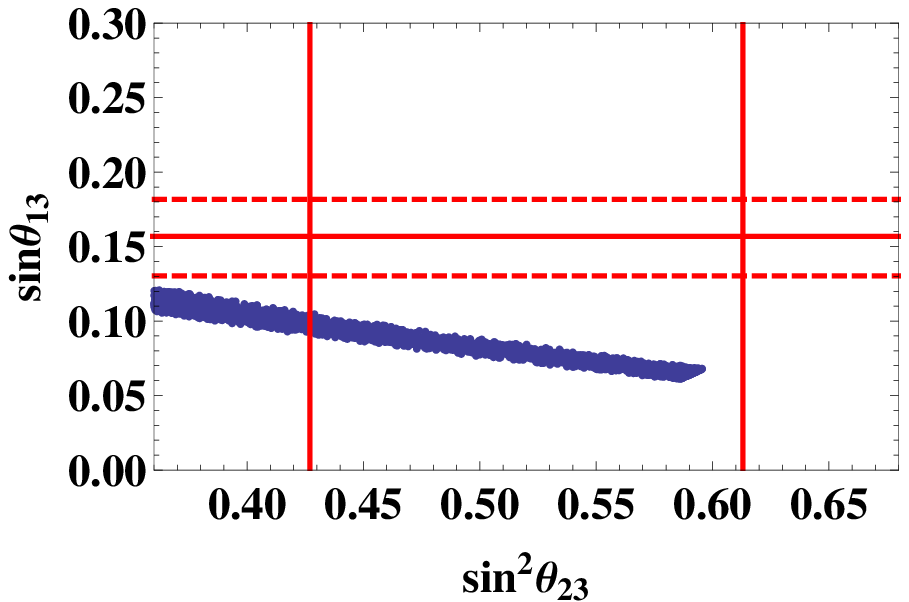} \\  
\\
\hspace{1.1cm}(c) & \hspace{1cm}(d) \\
\includegraphics[width=6.8cm]{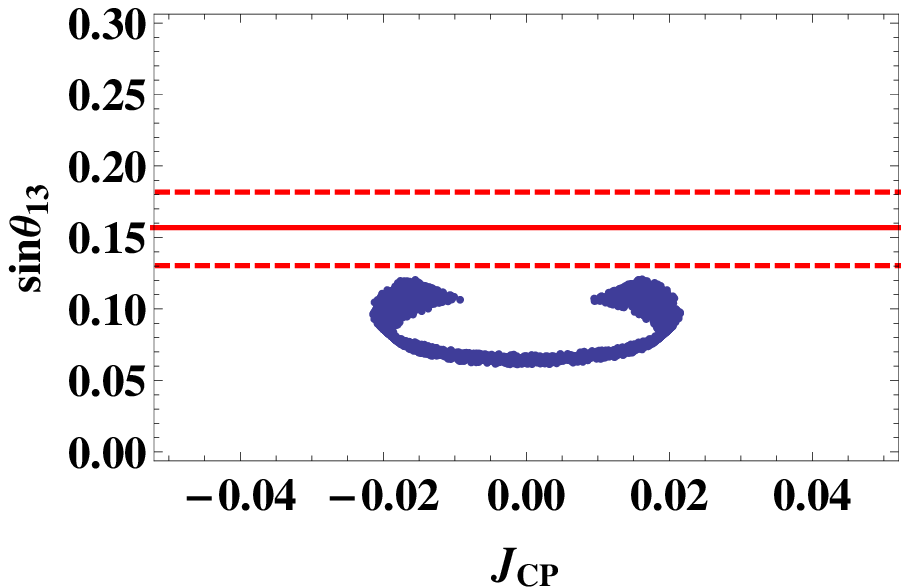} & \includegraphics[width=6.8cm]{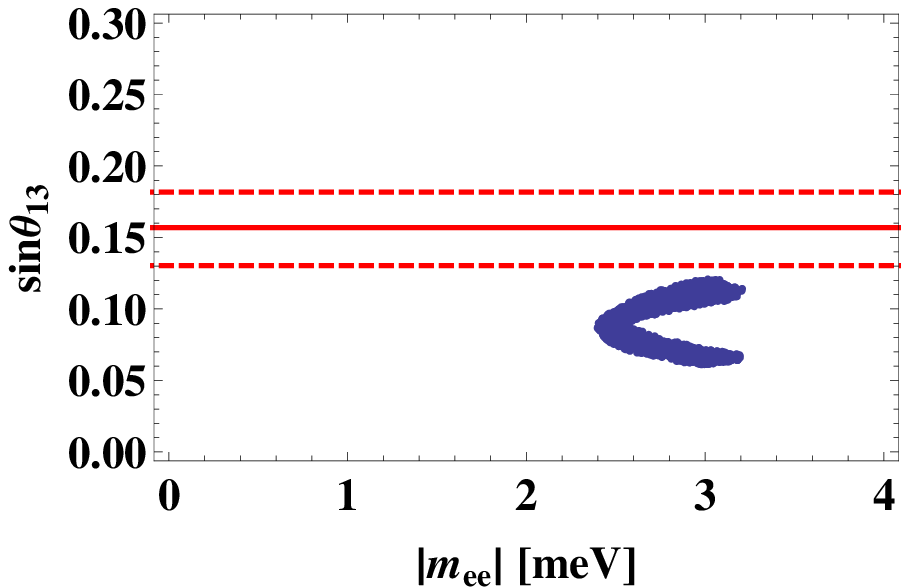} \\
\\
\multicolumn{2}{c}{\hspace{1cm}(e)}\\
\multicolumn{2}{c}{\includegraphics[width=6.8cm]{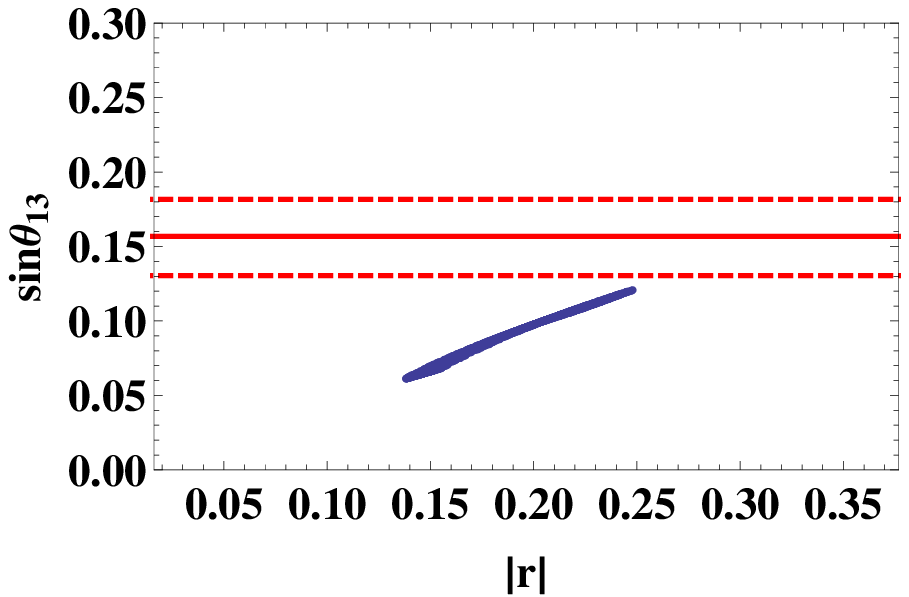}} \\
\\
\hline
\end{tabular}
\end{center}
\caption{Predicted regions of mixing angles, the Jarlskog invariant, the 
effective neutrino mass for the $0\nu \beta \beta $, and a favored region of $|r|$
in the case (i\hspace{-.1em}i) of NH with two zero texture. 
The notations in each figure are the same as ones in the Fig.~\ref{fig1}.}
\label{fig2}
\end{figure}
In these numerical calculations, we take the parameter $r$ to be complex as 
follows: 
\begin{equation}
r=|r|e^{i\alpha }, \qquad 
-\pi <\alpha <\pi .
\label{para}
\end{equation}
Our numerical results of mixing angles should be compared 
to the experimental data~\cite{Tortola:2012te} with $3\sigma $ as
\begin{align}
0.27\leq \sin ^2\theta _{12}
\leq 0.37,\quad 0.36\leq \sin ^2\theta _{23}
\leq 0.68,\quad 0.017\leq \sin ^2\theta _{13}
\leq 0.033.
\end{align}
All plots are 
within $3\sigma $ ranges of both $\Delta m_{\text{atm}}^2$ and 
$\Delta m_{\text{sol}}^2$~\cite{Tortola:2012te} as 
\begin{equation}
\Delta m_{\text{atm}}^2=(2.55_{-0.24}^{+0.19})\times 10^{-3}~\text{eV}^2,\qquad 
\Delta m_{\text{sol}}^2=(7.62_{-0.50}^{+0.58})\times 10^{-5}~\text{eV}^2.
\label{D}
\end{equation}
As seen from Figs.~\ref{fig1} (a) and \ref{fig1} (b), we find that the case (i) is marginal 
to reproduce the experimental results, and we predict, 
\begin{align}
&
0.320\lesssim \sin ^2\theta _{12}\lesssim 0.322,
\qquad 
0.660\lesssim \sin ^2\theta _{23}\lesssim 0.680,\qquad \nonumber \\
&
0.170~(0.130)\lesssim \sin ^2\theta _{13}~(\sin \theta _{13})\lesssim 0.196~(0.140).
\end{align}
And the Jarlskog invariant, which is one of parameters 
describing the size of CP-violation, is predicted as 
\begin{eqnarray}
|J_{CP}|\equiv|\mbox{Im}(U_{\alpha i}U_{\beta j}U_{\beta i}^\ast U_{\alpha j}^\ast)|\simeq 0.01,
\end{eqnarray}
within the $3\sigma $ range of the $\theta _{13}$ in the Fig.~\ref{fig1} (c). 
The non-vanishing $|J_{CP}|=\mathcal{O}(0.01)$ promises an observation of 
the leptonic CP-violation in the future long-baseline neutrino experiments. 
An effective mass for the neutrino-less double beta decay ($0\nu \beta \beta $); 
\begin{equation}
m_{ee}\equiv\sum _{i=1}^3m_iU_{ei}^2~, 
\end{equation}
is also predicted as 
\begin{equation}
|m_{ee}|\simeq 3.6~\text{meV},
\end{equation}
within the $3\sigma $ range of the $\theta _{13}$ in the Fig.~\ref{fig1} (d). 
The Heidelberg-Moscow experiment is currently giving the strongest bound on 
$|m_{ee}|$ as $|m_{ee}|\lesssim 210$ meV~\cite{KlapdorKleingrothaus:2000sn}. 
 KamLAND-Zen has reported the  bound  $|m_{ee}|=(260-540)$ 
meV~\cite{KamLANDZen:2012aa}, 
and will present the promising result in the near future.
The KamLAND-Zen and CUORE experiments are 
expected to reach $|m_{ee}|\simeq (20-80)$ meV~\cite{KamLANDZen:2012aa} and 
$|m_{ee}|=(24-93)$ meV~\cite{Arnaboldi:2002du}, respectively. 
Thus, it may be difficult to check this texture by the future experiments.

In these numerical calculations, we have scanned over a broader range of $|r|$ 
$(0.02\lesssim |r|\lesssim 0.4)$ than a range expected by the 
Eqs.~(\ref{mass-2texture1}) and (\ref{D}) within $3\sigma $ level to give 
complete predicted regions of physical quantities in the texture. It is 
actually found from the Fig.~\ref{fig1} (e) that a value around 
$r\sim \sqrt{\Delta m_{\text{sol}}^2/\Delta m_{\text{atm}}^2}$ 
is favored by the experimental data. 
\\

\noindent \underline{{\bf (i\hspace{-.1em}i) {\boldmath $b+c=0$ and $e=0$ case}}}
\\

The left-handed Majorana neutrino mass matrix is 
\begin{equation}
M_\nu =\frac{m_0^2}{m_R}V_{\text{TBM}}^T
\begin{pmatrix}
0 & 0 & 0 \\
0 & \frac{3}{4}f^2 & \frac{1}{2}\sqrt{\frac{3}{2}}f^2 \\
0 & \frac{1}{2}\sqrt{\frac{3}{2}}f^2 & 1+\frac{1}{2}f^2
\end{pmatrix}V_{\text{TBM}}.
\end{equation}
It can be also seen that the additional mixing, $\phi $, is too large 
to be consistent with the experimental data if  $f\sim\mathcal{O}(1)$, and thus
 we must take $f\ll 1$ as well as $e\ll 1$ in the previous case (i). 
The neutrino mass eigenvalues and the additional mixing angle are 
\begin{equation}
m_1=0,\quad \frac{m_2}{m_3}\simeq \frac{3}{4}f^2\equiv r,\quad 
\tan (2\phi )\simeq \sqrt{\frac{3}{2}}f^2.
\label{mass-2texture2}
\end{equation}
The relevant mixing matrix elements are written from the Eq.~(\ref{mass-2texture2}), 
\begin{equation}
\left |U_{e2}\right |\simeq \frac{1}{\sqrt{3}}\sqrt{1-\frac{2}{3}r^2},\quad 
\left |U_{e3}\right |\simeq \frac{\sqrt{2}}{3}r,\quad 
\left |U_{\mu 3}\right |\simeq 
\left |\frac{\sqrt{2}}{3}r+\frac{1}{\sqrt{2}}\sqrt{1-\frac{2}{3}r^2}\right |~.
\end{equation}
Results of numerical calculations for the mixing angles, 
the Jarlskog invariant, the effective mass, and the favored region of $|r|$ 
are shown in the Fig.~\ref{fig2}. We find that this case cannot explain the current experimental
 results because the value of the $\sin \theta _{13}$ are not within $3\sigma $ 
range on the $\sin ^2\theta _{23}$--$\sin \theta _{13}$ plane which is contrast to 
the case (i). \\

\noindent \underline{{\bf (i\hspace{-.1em}i\hspace{-.1em}i) {\boldmath $c=0$ and $e=0$ case}}}
\\

The left-handed Majorana neutrino mass matrix is 
\begin{equation}
M_\nu =\frac{m_0^2}{m_R}V_{\text{TBM}}^T
\begin{pmatrix}
0 & 0 & 0 \\
0 & \frac{3}{4}(f^2+1) & \frac{1}{2}\sqrt{\frac{3}{2}}(f^2-1) \\
0 & \frac{1}{2}\sqrt{\frac{3}{2}}(f^2-1) & \frac{1}{2}(f^2+1)
\end{pmatrix}V_{\text{TBM}}.
\end{equation}
Since the neutrino mass eigenvalues can be obtained by 
\begin{equation}
m_1=0,\qquad \frac{m_2}{m_3}=\frac{5+5f^2-\sqrt{25-46f^2+25f^4}}
{5+5f^2+\sqrt{25-46f^2+25f^4}}\equiv r,
\end{equation}
the parameter $f$ is evaluated as 
\begin{equation}
f^2\simeq \frac{25}{24}r\quad \text{or}\quad \frac{24}{25r}. 
\end{equation}
The additional mixing angle is 
\begin{equation}
\tan (2\phi )=\frac{2\sqrt{6}(1-f^2)}{1+f^2}\simeq -2\sqrt{6}\quad \text{or}\quad 2\sqrt{6},
\end{equation}
and thus it is conflict with the experimental results because of the large 
additional rotation. 

We conclude in this subsection that the case (i) is marginal to explain the 
experimental results but the cases of (i\hspace{-.1em}i) and 
(i\hspace{-.1em}i\hspace{-.1em}i) are excluded. 

\subsubsection{One zero texture}
Next, we discuss one zero textures leading to the additional 2-3 rotation to 
the TBM. There are three possible patterns of one zero texture as follows: 
\begin{equation}
\mbox{(I) $b+c=0$,~~~(I\hspace{-.1em}I) $c=0$,~~~ 
(I\hspace{-.1em}I\hspace{-.1em}I) $b=0$,} 
\end{equation}
in the Eq.~(\ref{general-texture}). Corresponding Dirac neutrino mass matrices 
can be obtained as 
\begin{equation}
\frac{M_D^{3\times 2}}{m_0}=\left\{
\begin{array}{cl}
\begin{pmatrix}
0 & \frac{e+f}{2} \\
\frac{1}{\sqrt{2}} & e \\
-\frac{1}{\sqrt{2}} & f
\end{pmatrix} & \mbox{ for (I) $b+c=0$}\\ 
\begin{pmatrix}
\frac{1}{2} & \frac{e+f}{2} \\
1 & e \\
0 & f
\end{pmatrix} & \mbox{ for (I\hspace{-.1em}I) $c=0$}\\
\begin{pmatrix}
\frac{1}{2} & \frac{e+f}{2} \\
0 & e \\
1 & f
\end{pmatrix} & \mbox{ for (I\hspace{-.1em}I\hspace{-.1em}I) $b=0$}
\end{array}
\right .,\label{MD1}
\end{equation}
after rescaling in a similar way to the cases of two zero textures. However, in
 fact, all these patterns lead to exactly the same predictions for the mixing 
angles as we will show later. Therefore, we focus only on the texture of the 
case (I) $b+c=0$ in detail. 

Then, the right-handed Majorana and the Dirac neutrino mass matrices are 
written as 
\begin{equation}
M_R^{2\times 2}= m_R
\begin{pmatrix}
1 & 0 \\
0 & 1
\end{pmatrix},\qquad \frac{M_D^{3\times 2}}{m_0}=
\begin{pmatrix}
0 & \frac{e+f}{2} \\
\frac{1}{\sqrt{2}} & e \\
-\frac{1}{\sqrt{2}} & f
\end{pmatrix},
\label{MajoranaDirac}
\end{equation}
respectively. By using the seesaw mechanism, the left-handed Majorana neutrino 
mass matrix is 
\begin{equation}
M_{\nu }=\frac{m_0^2}{m_R}
\begin{pmatrix}
\frac{1}{4}(e+f)^2 & \frac{1}{2}e(e+f) & \frac{1}{2}(e+f)f \\
\frac{1}{2}e(e+f) & \frac{1}{2}+e^2 & -\frac{1}{2}+ef \\
\frac{1}{2}(e+f)f & -\frac{1}{2}+ef & \frac{1}{2}+f^2
\end{pmatrix}.
\label{LHN-matrix}
\end{equation}
In this case, the left-handed Majorana neutrino mass matrix is as follow: 
\begin{equation}
M_{\nu }=\frac{m_0^2}{m_R}\ V_{\text{TBM}}^T 
\begin{pmatrix}
0 & 0 & 0 \\
0 & \frac{3}{4}(e+f)^2 & -\frac{1}{2}\sqrt{\frac{3}{2}}(e-f)(e+f) \\
0 & -\frac{1}{2}\sqrt{\frac{3}{2}}(e-f)(e+f) & 1+\frac{1}{2}(e-f)^2
\end{pmatrix}
V_{\text{TBM}},
\end{equation}
where we take the rescaled right-handed Majorana neutrino mass matrix as one 
given in the Eq.~(\ref{general-texture}). 
When we take $e,f\ll 1$ to obtain appropriate size
 of the additional rotation, the neutrino mass eigenvalues are approximated by 
\begin{equation}
m_1=0,\qquad \frac{m_2}{m_3}\simeq \frac{3}{4}(e+f)^2\equiv r.
\label{mass-eigenvalues}
\end{equation}
As the result, the angle of the additional rotation is described by 
\begin{equation}
\tan (2\phi )\simeq -\sqrt{\frac{3}{2}}(e-f)(e+f)\equiv \sqrt{6}\lambda .
\label{mixing-deviation-TBM}
\end{equation}
Here it might be convenient for the following discussion to define new 
parameter, $\lambda $, as $\lambda \equiv \tan(2\phi )/\sqrt{6}$. 
In this case, relevant mixing matrix elements are given as 
\begin{align}
|U_{e2}|\simeq \sqrt{\frac{1}{3}-\frac{\lambda ^2}{2}},\quad 
|U_{e3}|\simeq \left |\frac{\lambda }{\sqrt{2}}\right |, \quad 
|U_{\mu 3}|\simeq \left |\frac{\lambda }{\sqrt{2}}+\sqrt{\frac{1}{2}-\frac{3\lambda ^2}{4}}\right |.
\end{align}
Now, the parameters $e$ and $f$ can be reparameterized as 
\begin{equation}
e=\frac{2re^{i\alpha }-3\lambda e^{i\beta }}{2\sqrt{3re^{i\alpha }}},\qquad 
f=\frac{2re^{i\alpha }+3\lambda e^{i\beta }}{2\sqrt{3re^{i\alpha }}},
\label{repa}
\end{equation}
including phases with the Eq.~(\ref{para}) and 
\begin{equation}
-\pi <\beta <\pi,
\label{para1}
\end{equation}
where we generically take $e$ and $f$ as complex. The parameters $r$ and 
$\lambda $ are real. We show the results of numerical calculation within the 
regions of parameters, the Eqs.~(\ref{para}), (\ref{para1}), and $|\lambda |\leq 0.4$. 
Since it is clear that relatively large value of 
$|\lambda |$, which determines the size of additional rotation, cannot be 
allowed by the experimental data, we scanned within a range of small values of 
$|\lambda |$. 
\begin{figure}
\begin{center}
\begin{tabular}{cc}
\hline
\multicolumn{2}{c}{One zero texture in the NH} \\
\hline \hline
\multicolumn{2}{c}{$|\lambda |\leq 0.4$} \\
\hline \\
\hspace{1.1cm}(a) & \hspace{1cm}(b) \\
\includegraphics[width=6.8cm]{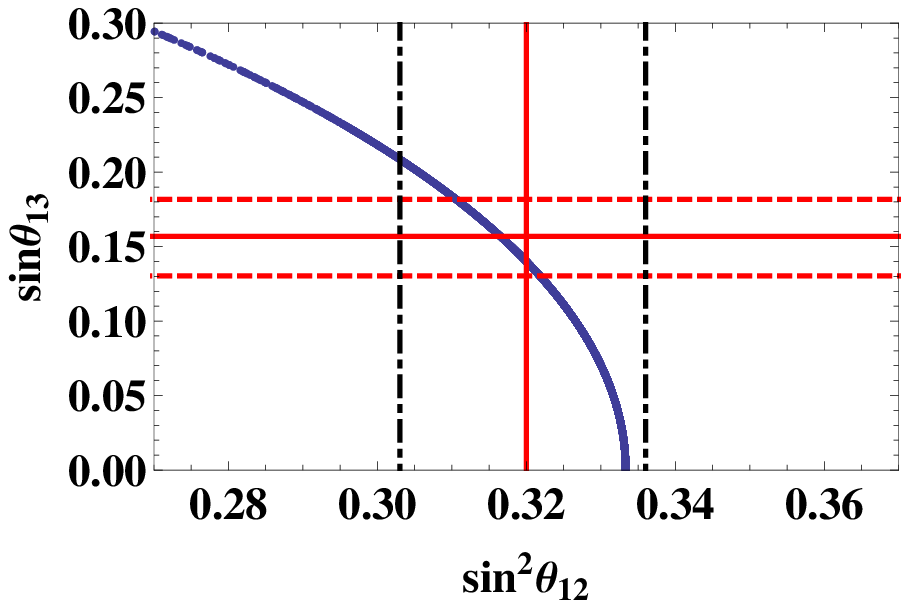} & \includegraphics[width=6.8cm]{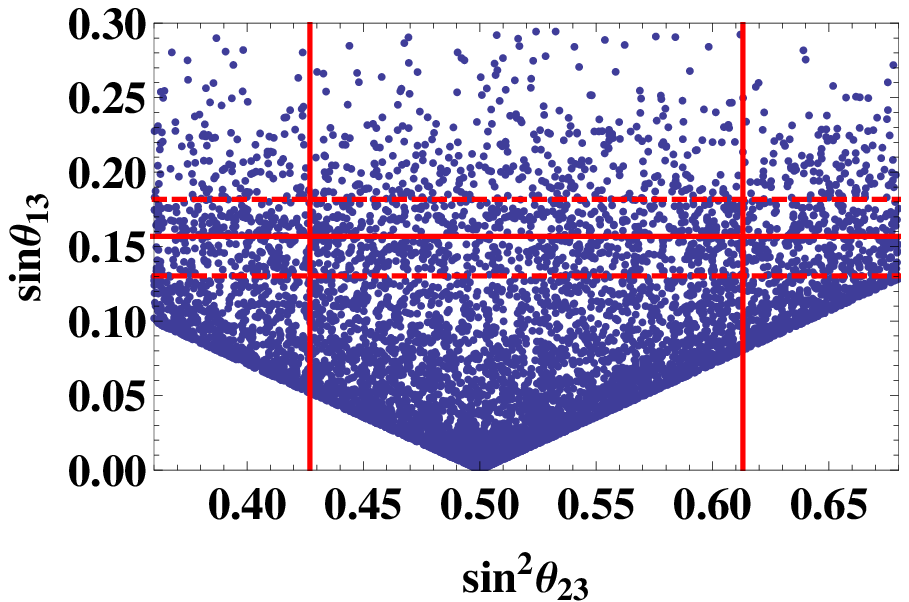} \\  
\\
\hspace{1.1cm}(c) & \hspace{1cm}(d) \\
\includegraphics[width=6.8cm]{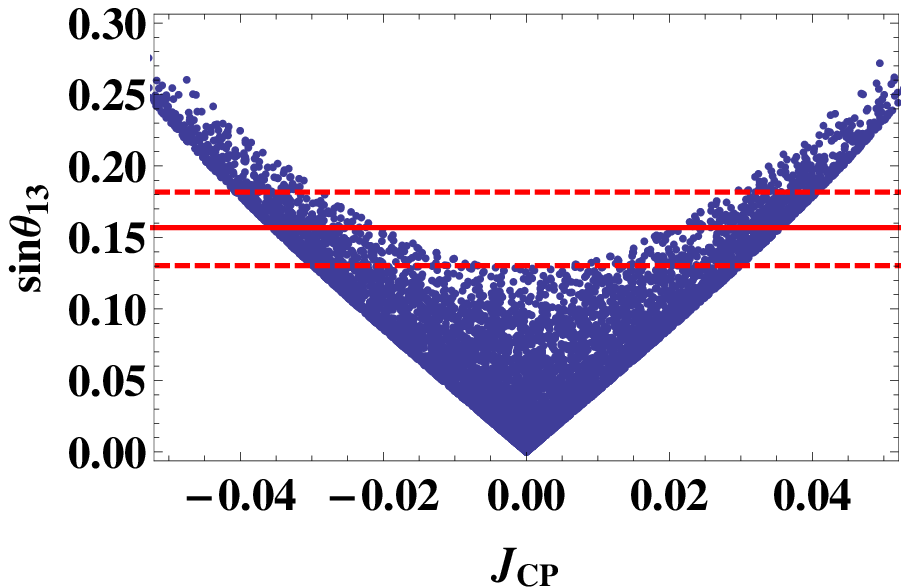} & \includegraphics[width=6.8cm]{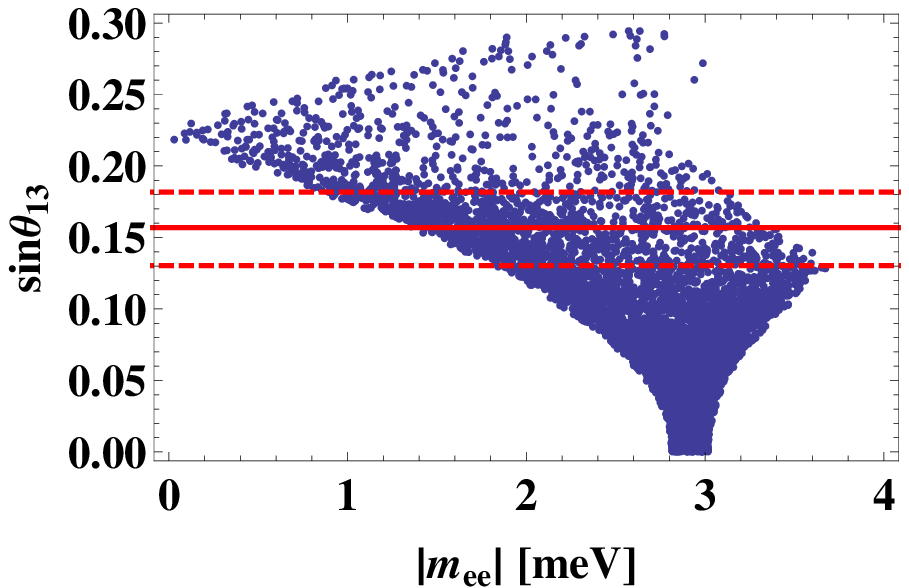} \\
\\
\multicolumn{2}{c}{\hspace{1cm}(e)}\\
\multicolumn{2}{c}{\includegraphics[width=6.8cm,bb=0 0 260 170]{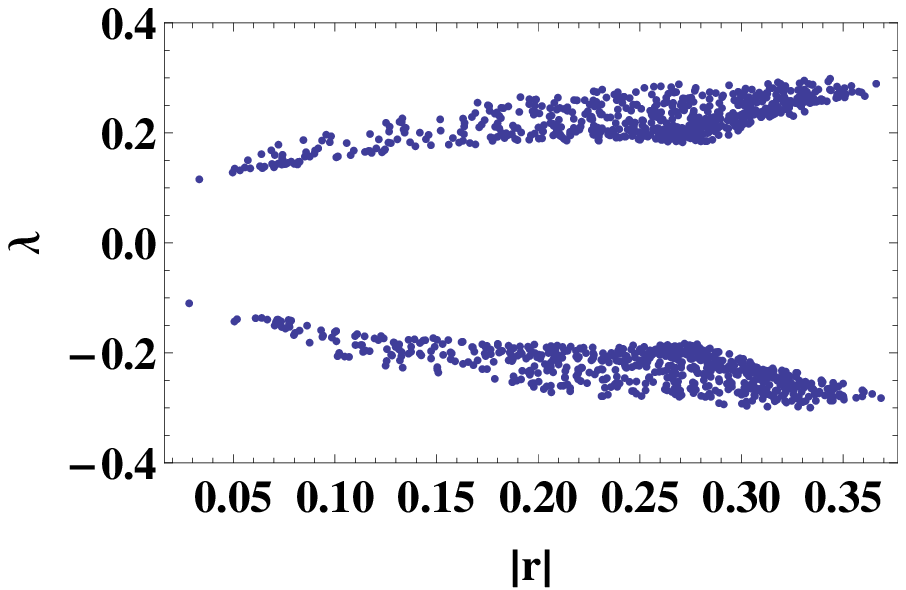}} \\
\\
\hline
\end{tabular}
\end{center}
\caption{Predicted regions of mixing angles, the Jarlskog invariant, the 
effective neutrino mass for the $0\nu \beta \beta $, and a favored region of $|r|$ and $\lambda $ 
in the one zero texture of NH. The other notations of each figure are the same as ones 
in the Fig.~\ref{fig1}. The figure (e) is a plot on $|r|$--$\lambda $ plane where all points 
are within $3\sigma $ range of three mixing angles.}
\label{fig3}
\end{figure}
As seen in the Fig.~\ref{fig3} (a), there is a clear correlation between $\theta _{12}$ and 
$\theta _{13}$. We predict 
\begin{equation}
0.310\lesssim \sin ^2 \theta _{12}\lesssim 0.322,
\end{equation}
within $3\sigma $ range of the $\sin\theta _{13}$, which is expected to be 
testable in the future  precise  neutrino experiments. On the other hand, there
 is no clear correlation between $\theta _{23}$ and $\theta _{13}$ 
because of the presence of two unknown phases ($\alpha $ and $\beta $) and 
complicated $\lambda $ dependence of $|U_{\mu3}|$ compared with $|U_{e2}|$ in 
the Fig.~\ref{fig3} (b). In the Figs.~\ref{fig3} (c) and~\ref{fig3} (d), 
$|J_{CP}|$ and $|m_{ee}|$ are also predicted as 
\begin{equation}
|J_{CP}|\lesssim 0.04,\qquad
0.8~\text{meV}\lesssim |m_{ee}|\lesssim 3.6~\text{meV},
\end{equation}
within $3\sigma $ range of the $\theta _{13}$. A favored region of the parameters
 $|r|$ and $\lambda $ is also shown in the Fig.~\ref{fig3} (e) where all points 
are within $3\sigma $ range of three mixing angles. 

For the other possible cases of the one zero textures, (I\hspace{-.1em}I) $c=0$ 
and (I\hspace{-.1em}I\hspace{-.1em}I) $b=0$, we can always take the same 
definitions of $r$ and $\lambda $ as ones in the case (I) without the loss of 
generality although forms of function for $r$ and $\lambda $ in terms of $e$ and
 $f$ are different among all cases. This means that predicted 
regions of the mixing angles, the Jarlskog invariant, and the effective 
neutrino mass for the $0\nu \beta \beta $ from three cases 
(I)-(I\hspace{-.1em}I\hspace{-.1em}I) are the same. Note that there are two 
free parameters $r$ and $\lambda $ to determine the neutrino mass ratio and the 
size of additional mixing angles in the one zero textures in contrast to the 
cases of the two zero textures, where each model described by only $r$, i.e. the
 additional mixing angles are tightly related with the neutrino mass ratio. 

\subsection{Inverted neutrino mass hierarchy}
Let us discuss the case of the inverted neutrino mass hierarchy (IH). 
In order to lead to the additional 2-3 rotation, the $(1,2)$, $(1,3)$, $(2,1)$,
 and $(3,1)$ elements in the Eq.~(\ref{left-handed-Majorana-TBM}) should vanish. 
These conditions are given as 
\begin{equation}
A=a+b+c=0,\quad C=c-b,\quad D=2d-e-f=0. 
\end{equation}
Then, the left-handed Majorana neutrino mass matrix is written by 
\begin{equation}
M_\nu =\frac{v^2}{m_R}V_{\text{TBM}}^T
\begin{pmatrix}
6b^2 p & 0 & 0 \\
0 & \frac{3}{4}(e+f)^2 & -\frac{1}{2}\sqrt{\frac{3}{2}}(e-f)(e+f) \\
0 & -\frac{1}{2}\sqrt{\frac{3}{2}}(e-f)(e+f) & \frac{1}{2}(e-f)^2
\end{pmatrix}V_{\text{TBM}}.
\label{left-handed-Majorana-general-inverted}
\end{equation}
By reparameterizing of the matrix elements, we have 
\begin{equation}
M_R^{2\times 2}=m_R
\begin{pmatrix}
1 & 0 \\
0 & 1
\end{pmatrix}, \quad \quad \frac{M_D^{3\times 2}}{m_0}=
\begin{pmatrix}
-2 & \frac{e+f}{2} \\
1 & e \\
1 & f
\end{pmatrix}.
\label{general-texture-inverted}
\end{equation}
The mass eigenvalues and the additional mixing angle are given by 
\begin{equation}
m_3=0,\quad \frac{m_2}{m_1}=\frac{1}{24}\left (5e^2+2ef+5f^2\right )\equiv r',\quad 
\tan \phi =\sqrt{\frac{2}{3}}\ \frac{e-f}{e+f}~,
\label{rin}
\end{equation}
where $e$ and $f$ are supposed to be real, for simplicity, and the definition of $r'$ 
differs from $r$ in the NH case. Here note that if we consider one zero textures 
($e=0$, $f=0$, or $e+f=0$), the additional mixing 
becomes too large to be consistent with the experimental results. Therefore, we
 cannot consider zero textures for the Dirac neutrino mass matrix 
in the Eq.~(\ref{general-texture-inverted}), and the texture of the 
Eq.~(\ref{general-texture-inverted}) is a minimal one to lead to the additional
 2-3 rotation to the TBM in the IH case. 

\begin{figure}
\begin{center}
\begin{tabular}{cc}
\hline
\multicolumn{2}{c}{Minimal texture in the IH} \\
\hline \hline
\multicolumn{2}{c}{$|\lambda |\leq 0.4$} \\
\hline \\
\hspace{1.1cm}(a) & \hspace{1cm}(b) \\
\includegraphics[width=6.8cm]{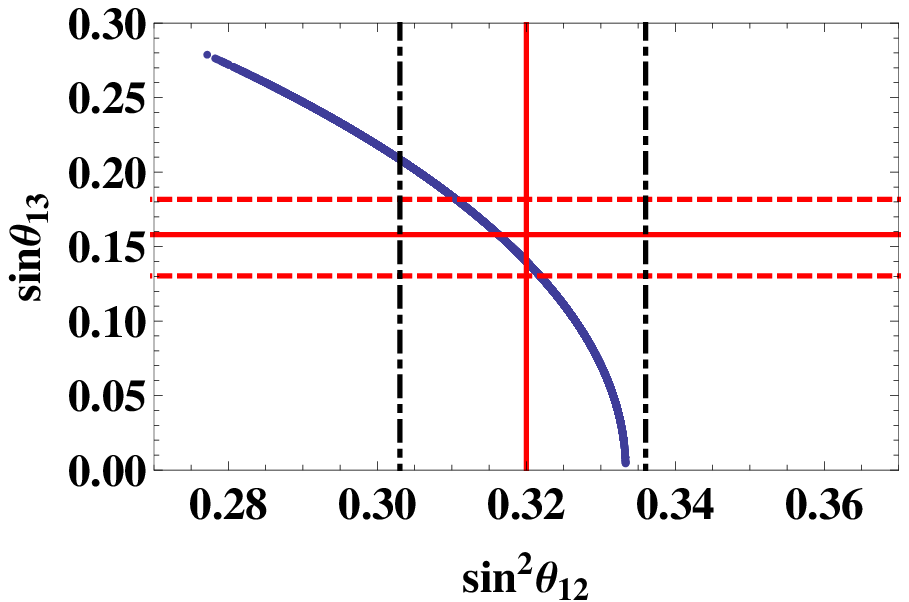} & \includegraphics[width=6.8cm]{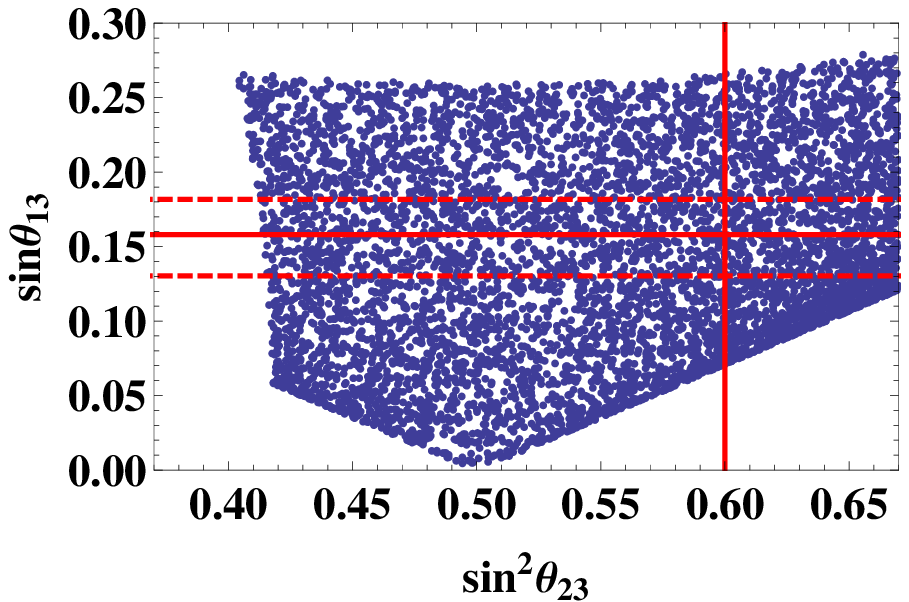} \\  
\\
\hspace{1.1cm}(c) & \hspace{1cm}(d) \\
\includegraphics[width=6.8cm]{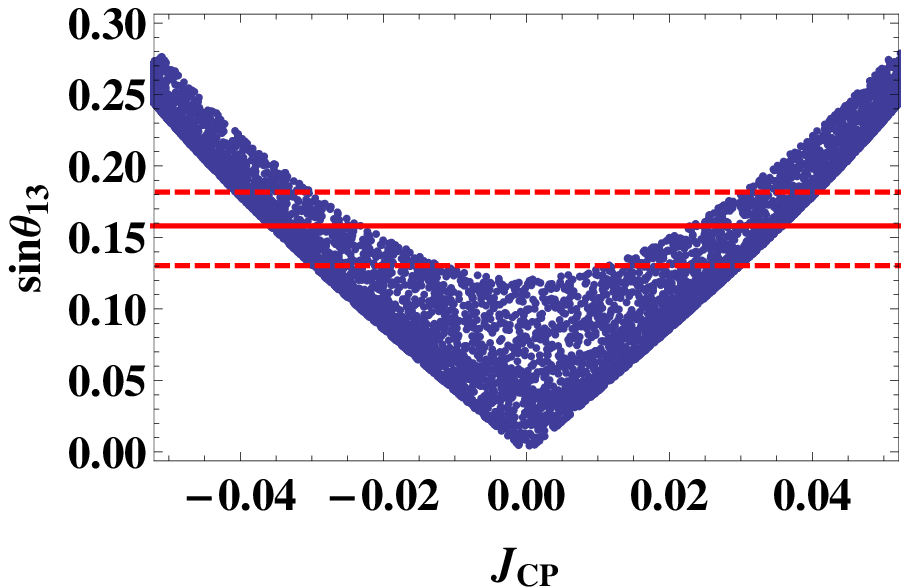} & \includegraphics[width=6.8cm]{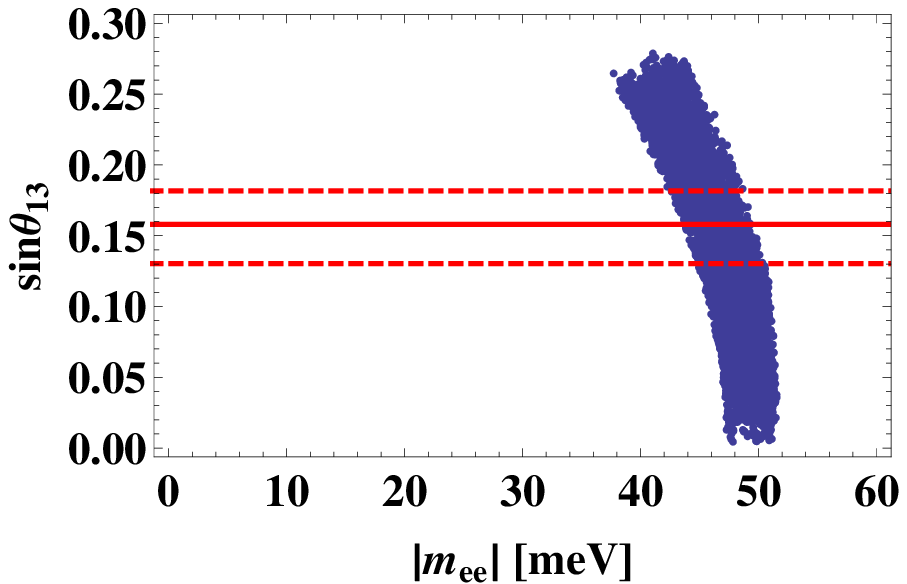} \\
\\
\multicolumn{2}{c}{\hspace{1cm}(e)}\\
\multicolumn{2}{c}{\includegraphics[width=6.8cm]{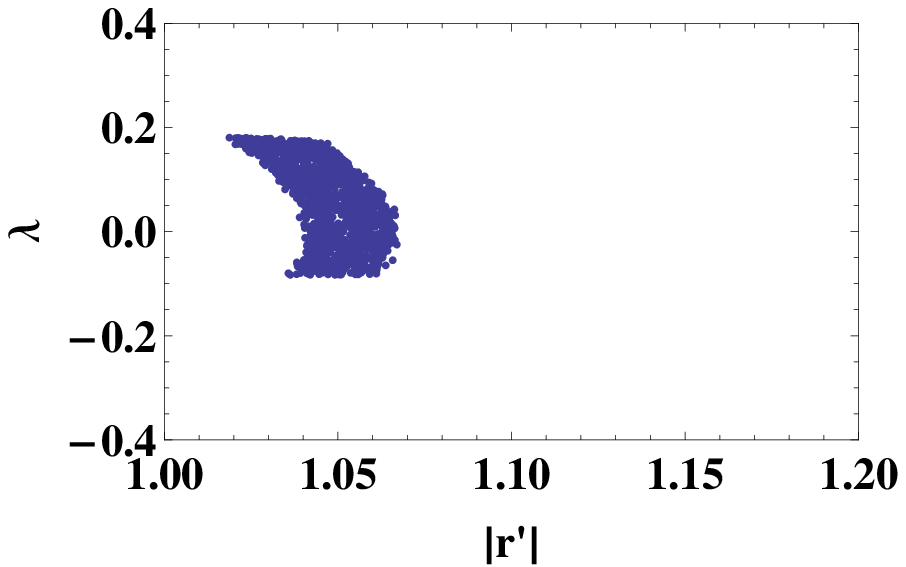}} \\
\\
\hline
\end{tabular}
\end{center}
\caption{Predicted regions of mixing angles, the Jarlskog invariant, the 
effective neutrino mass for the $0\nu \beta \beta $, and a favored region of $|r|$
 and $\lambda $ in the minimal texture of the IH. The notations of each figure 
are the same as ones in the Fig.~\ref{fig1}. The figure (e) is a plot on 
$|r'|$--$\lambda $ plane where all points are within $3\sigma $ range of three 
mixing angles.}
\label{fig4}
\end{figure}

In our calculations, we take $e$ and $f$ to be complex, and 
they are generically described by one complex number $r'$~(with $|r'|>1$) and 
one real number $\lambda ~(\equiv \tan (2\phi )/\sqrt{6})$ with 
phase $\beta $ as $\lambda e^{i\beta }$. The results of numerical calculations 
are shown in the Fig.~\ref{fig4}. All plots are within $3\sigma $ ranges 
of the experimental data~\cite{Tortola:2012te} for the IH case as 
\begin{align}
\Delta m_{\text{atm}}^2=-(2.43_{-0.22}^{+0.21})\times 10^{-3}~\text{eV}^2, \qquad 
&\Delta m_{\text{sol}}^2=(7.62_{-0.50}^{+0.58})\times 10^{-5}~\text{eV}^2, \nonumber 
\label{Din}
\end{align}
and should be also compared with the experimental data as 
\begin{align}
0.27\leq \sin ^2\theta _{12}\leq 0.37,\quad 
0.37\leq \sin ^2\theta _{23}&\leq 0.67,\quad 
0.017\leq \sin ^2\theta _{13}\leq 0.033.
\end{align}
We find that there is also a clear correlation between $\theta _{12}$ and 
$\theta _{13}$ shown in the Fig.~\ref{fig4}~(a) as well as the NH case, and we predict 
\begin{equation}
0.310\lesssim \sin ^2\theta _{12}\lesssim 0.322,
\end{equation}
within $3\sigma $ range of the $\sin\theta _{13}$. Further, it is also seen that 
there are some bounds, which come from the neutrino mass ratio, for three 
mixing angles as 
\begin{equation}
0.276\lesssim \sin ^2\theta_{12},\qquad 
0.400\lesssim \sin ^2\theta_{23},\qquad 
\sin ^2\theta _{13}~(\sin \theta _{13})\lesssim 0.0784~(0.280),
\end{equation}
in the contrast to the case (I) of NH with one zero texture shown in the Fig.~\ref{fig3}. 
We also predict 
\begin{equation}
0.01\lesssim |J_{CP}|\lesssim 0.04,\qquad
42~\text{meV}\lesssim |m_{ee}|\lesssim 51~\text{meV},
\end{equation}
within $3\sigma$ range of the $\theta_{13}$ in the Figs.~\ref{fig4} (c) and 
\ref{fig4} (d). These predictions are expected to be testable in the future 
precise neutrino experiments. Especially, the effective mass  in this IH case 
might be 
measured by the future KamLAND-Zen and CUORE experiments. 

Also in these numerical calculations, we have scanned over a broader range of 
$|r'|$~($1<|r'|\lesssim 1.2$) than a range expected by the Eqs.~(\ref{rin}) and 
(\ref{Din}) within $3\sigma $ level to present complete predicted regions of 
physical quantities from this texture. It is also seen 
from the Fig.~\ref{fig4} (e) that a value around 
$r'\sim \sqrt{(\Delta m_{\text{atm}}^2+\Delta m_{\text{sol}}^2)/\Delta m_{\text{atm}}^2}$ 
is favored by the experimental results. 

\section{Texture with Cabibbo angle and neutrino mass ratio}

In the previous sections, we have presented relatively general discussions of 
$3\times 2$ textures for the Dirac neutrino mass matrix (with zeros) in the context of 
(2+1) seesaw mechanism. At the end of this work, we consider interesting 
textures with specific assumptions and show their predictions. A basic strategy
 to make assumptions is to regard some small parameters in our context as the 
Cabibbo angle or the neutrino mass ratio. The reasons for such identification 
are just that it is now phenomenologically allowed in the current stage of 
neutrino oscillation experiments. We will not construct some actual high energy
 models to make the assumption valid in this work but such kind of assumptions 
might well motivate for studies of theory and/or symmetry in the quark/lepton 
sectors behind the SM. Here we still concentrate on texture analyses with the 
above assumptions.

In the previous sections, we investigate two and one zero textures. In the two zero
 textures, the model is described by one parameter $r$ after rescaling the 
right-handed Majorana neutrino mass matrix. Then, the parameter $r$ is approximated by 
the ratio between two mass squared differences of neutrino and determines the 
size of additional 2-3 mixing angle $\phi $. As the results we predict some 
clearly correlated regions of the mixing angles. 
On the other hand, in the one zero textures, we have one more 
parameter $\lambda $ in addition to $r$, which characterize the models. The 
$\lambda $ determines the magnitude of the additional mixing angle and 
it is restricted to be small by the current experimental data. 
In this section, we consider the model of one zero texture 
with the assumptions that the parameter $\lambda $ is identified with the 
Cabibbo angle or the related parameter with the neutrino mass ratio. 

\subsection{One zero texture with the Cabibbo angle}

\begin{figure}
\begin{center}
\begin{tabular}{cc}
\hline
\multicolumn{2}{c}{One zero texture with the Cabibbo angle 
and neutrino mass ratio in the NH} \\
\hline \hline
\multicolumn{2}{c}{$\lambda=0.225$}\\
\hline \\
\hspace{1.1cm}(a) & \hspace{1cm}(b) \\
\includegraphics[width=6.8cm]{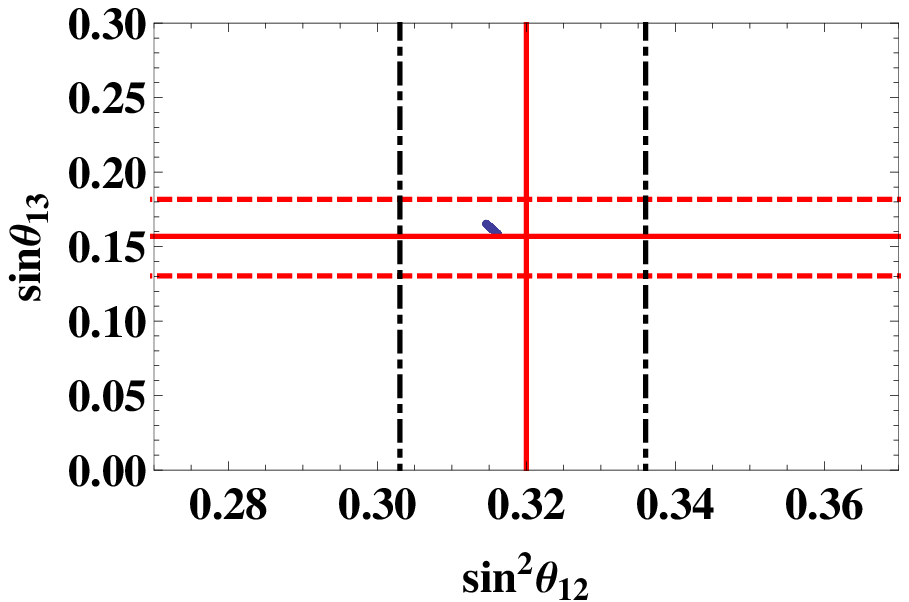} & \includegraphics[width=6.8cm]{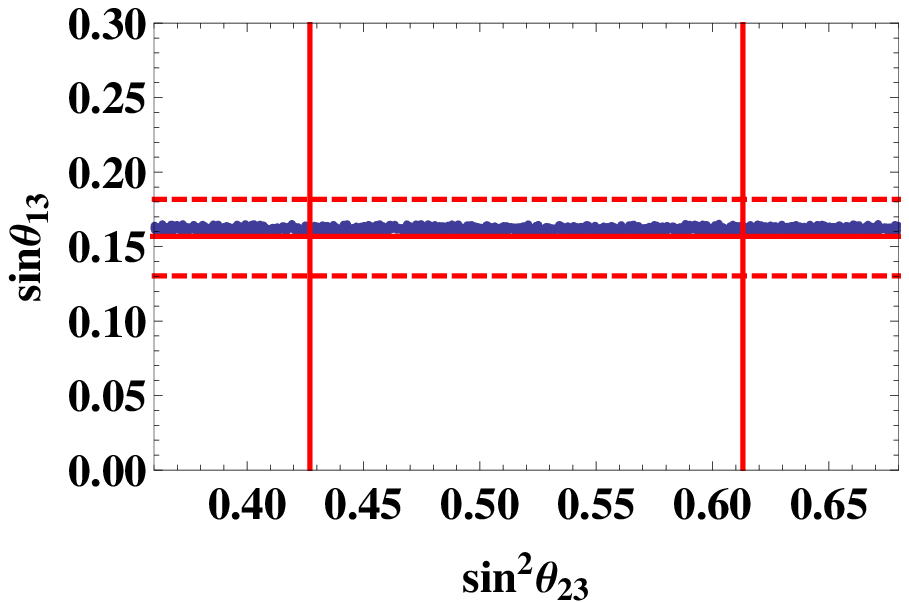} \\
\\
\hspace{1.1cm}(c) & \hspace{1cm}(d) \\
\includegraphics[width=6.8cm]{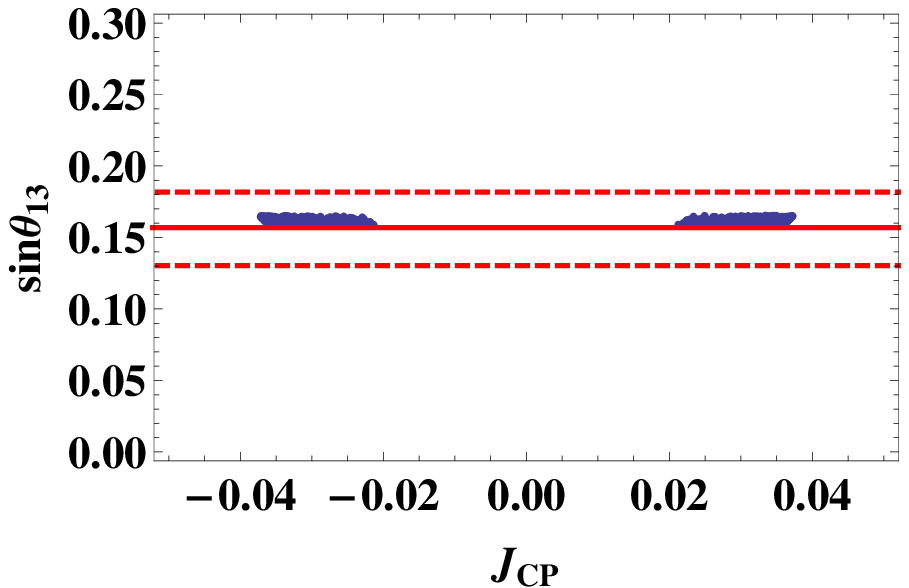} & \includegraphics[width=6.8cm]{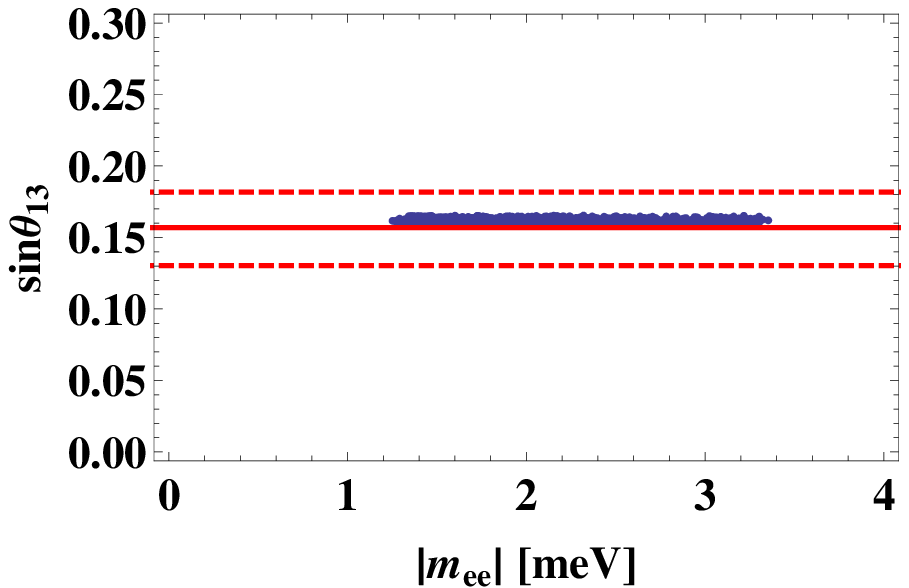} \\
\\
\multicolumn{2}{c}{\hspace{1cm}(e)}\\
\multicolumn{2}{c}{\includegraphics[width=6.8cm]{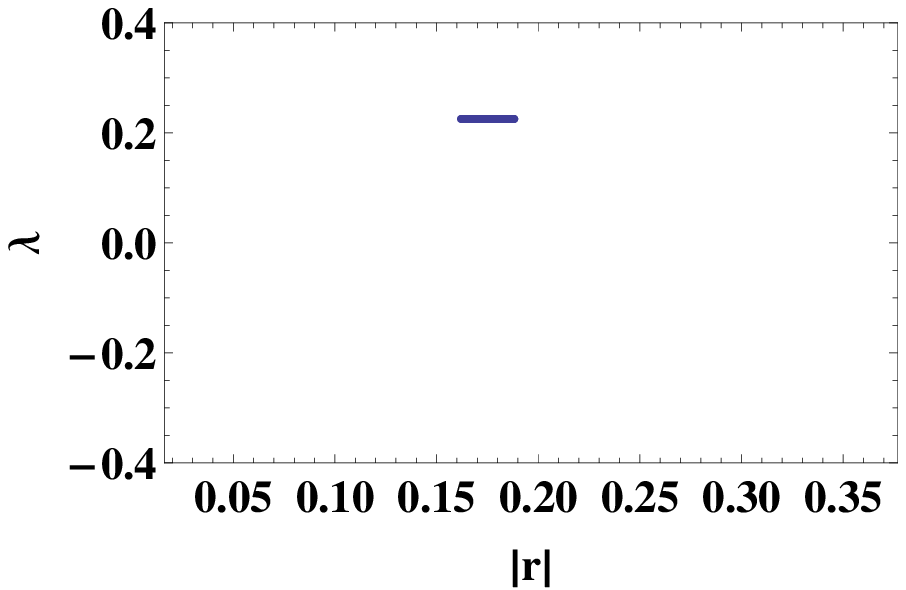}} \\
\\
\hline
\end{tabular}
\end{center}
\caption{Predicted regions of mixing angles, the Jarlskog invariant, the 
effective neutrino mass for the $0\nu\beta\beta$, and a favored region of $|r|$
 and $\lambda$ in the one zero texture with $\lambda =0.225$. 
The notations of each figure are the same as ones in the Fig.~\ref{fig1}. 
The figure (e) is a plot on $|r|$--$\lambda$ plane where all points are within 
$3\sigma $ range of three mixing angles.}
\label{fig5}
\end{figure}

First, we discuss the case that the model parameter $\lambda$ in the 
Eq.~(\ref{mixing-deviation-TBM}) is taken as the Cabibbo angle, 
$\lambda=0.225$. The numerical calculation of the case is given in the Fig.~\ref{fig5} 
and should be compared with the case in the Fig.~\ref{fig3}. 
\begin{figure}
\begin{center}
\begin{tabular}{cc}
\hline
\multicolumn{2}{c}{One zero texture with the Cabibbo angle and neutrino mass ratio 
in the NH}\\
\hline \hline
\multicolumn{2}{c}{$\lambda /\sqrt{2}=r$}\\
\hline \\
\hspace{1.1cm}(a) & \hspace{1cm}(b) \\
\includegraphics[width=6.8cm]{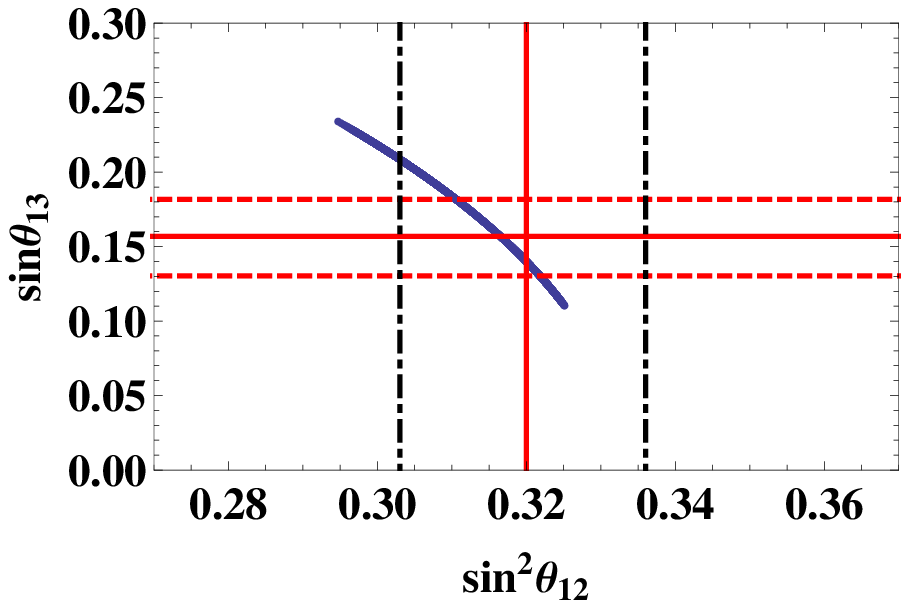} & \includegraphics[width=6.8cm]{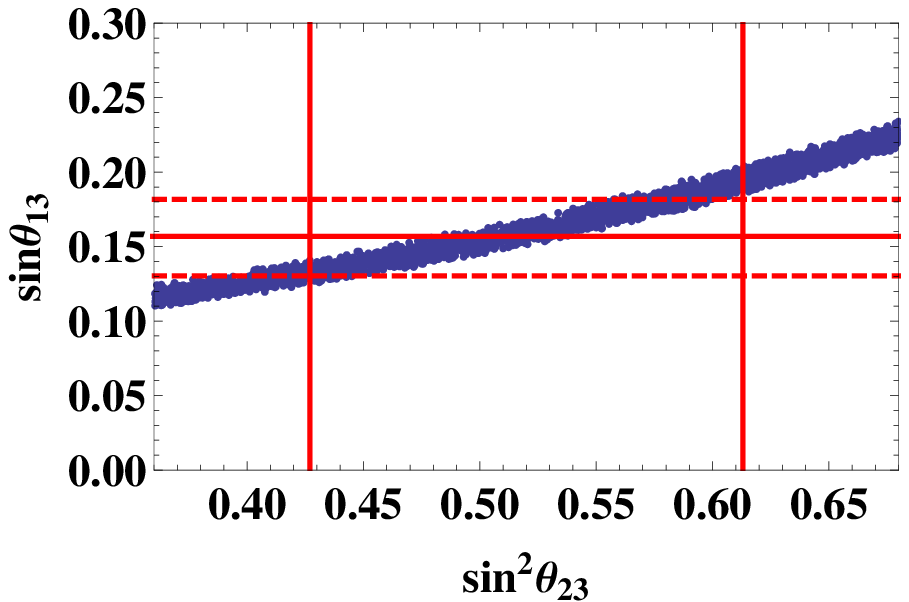}\\
\\
\hspace{1.1cm}(c) & \hspace{1cm}(d) \\
\includegraphics[width=6.8cm]{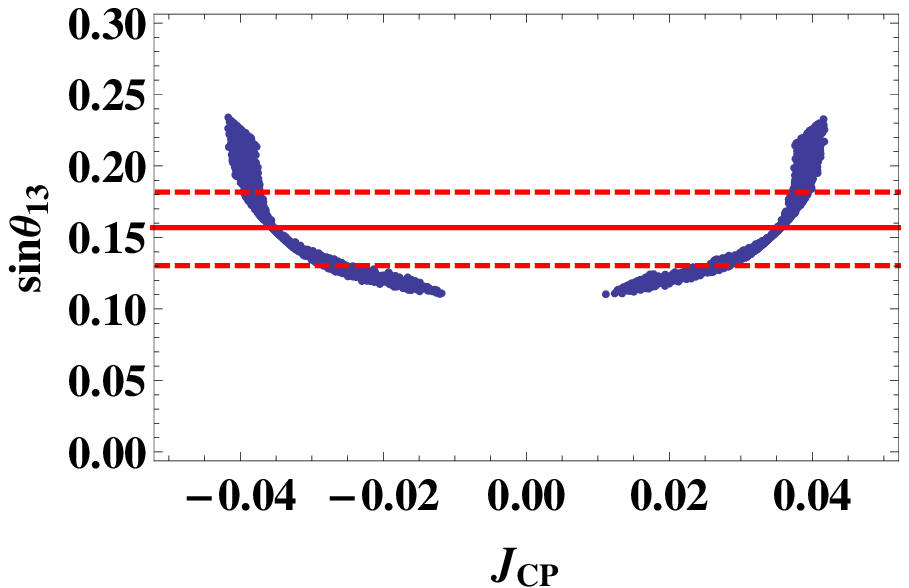} & \includegraphics[width=6.8cm]{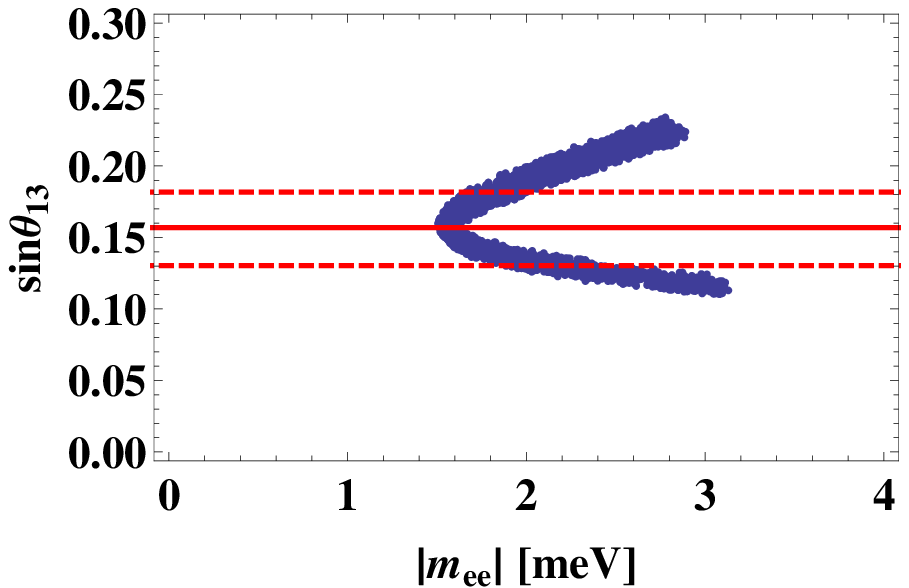}\\
\\
\multicolumn{2}{c}{\hspace{1cm}(e)}\\
\multicolumn{2}{c}{\includegraphics[width=6.8cm]{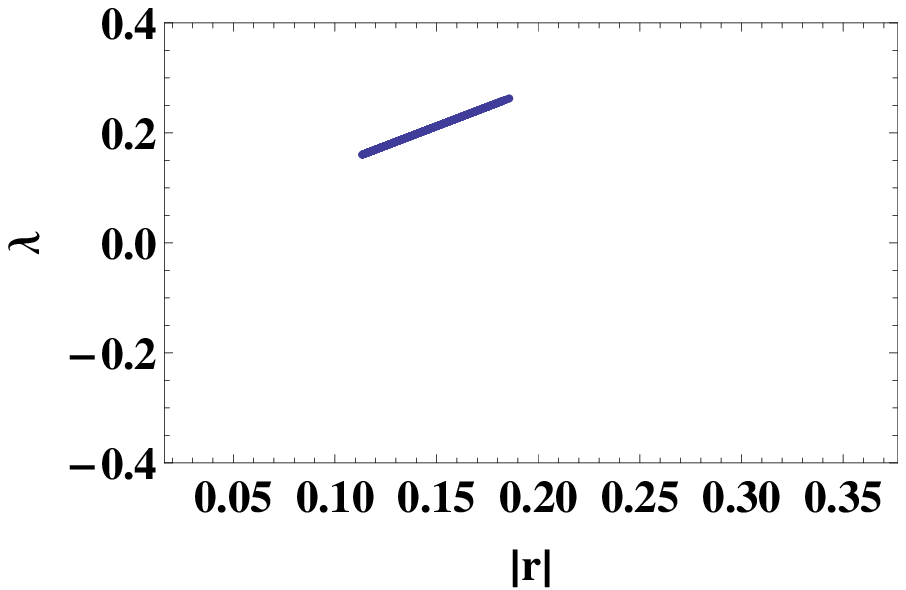}} \\
\\
\hline
\end{tabular}
\end{center}
\caption{Predicted regions of mixing angles, the Jarlskog invariant, the 
effective neutrino mass for the $0\nu \beta \beta $, and a favored region of $|r|$ and $\lambda $ 
in the one zero texture with $\lambda /\sqrt{2}=r$. The notations of each figure are 
the same as ones in the Fig.~\ref{fig1}. The figure (e) is a plot on $|r|$--$\lambda $ plane 
where all points are within $3\sigma $ range of three mixing angles.}
\label{fig6}
\end{figure}
The same numerical setup for the other parameter apart from $\lambda $ is taken 
as ones in the section 3.1.2. In the Fig.~\ref{fig5} (a), the $\sin ^2\theta _{12}$ 
and $\sin ^2\theta _{13}$~($\sin \theta _{13}$) are predicted as 
\begin{equation}
\sin ^2\theta _{12}\simeq 0.315,\qquad
\sin ^2\theta _{13}~(\sin \theta _{13})\simeq 0.0272~(0.165),
\end{equation}
which are close to the best fit values in the current experimental data. 
On the other hand, we cannot predict the value of the $\sin ^2\theta _{23}$ as 
seen the Fig.~\ref{fig5} (b). Regarding with the $J_{CP}$ and $|m_{ee}|$, they 
are predicted as 
\begin{equation}
0.02\lesssim |J_{CP}|\lesssim 0.04,\qquad 
1.2~\text{meV}\lesssim |m_{ee}|\lesssim 3.4~\text{meV},
\end{equation}
within $3\sigma $ range of the $\theta _{13}$ in the Figs.~\ref{fig5} (c) and 
\ref{fig5} (d). Such a large $J_{CP}$ is expected to be measured in the future 
long-baseline neutrino experiments. The Fig.~\ref{fig5} (e) is a plot on 
$|r|$--$\lambda $ plane where all points are within $3\sigma $ range of three 
mixing angles. This kind of model such that the additional rotation is 
described by the Cabibbo angle might be constructed by a high energy 
theory or symmetry for quark-lepton sectors. 

\subsection{One zero texture with the neutrino mass ratio}

The next concern is to investigate the one zero texture including only one small
 parameter as well as the two zero texture case, i.e. $\lambda $ is also related 
with the ratio between two mass squared differences of neutrinos 
because the ratio is naturally appeared in neutrino mass models. 

We take  $\lambda /\sqrt{2}=r$ in the Eq.~(\ref{repa}). In the case, the 
neutrino mass matrix is generally described one (complex) parameter, 
$r$.\footnote{The factor $\sqrt{2}$ in $\lambda /\sqrt{2}=r$ may also be 
interesting if $r$ and $\lambda$ are just taken as the neutrino mass ratio, 
$r=m_2/m_3=\sqrt{\Delta m_{\rm sol}^2/\Delta m_{\rm atm}^2}\simeq 0.16$, and as 
the Cabibbo angle, $\lambda /\sqrt{2}\simeq 0.225/\sqrt{2}\simeq 0.16$, respectively.} 
The results of numerical calculation are shown in the Fig.~\ref{fig6} 
with $\lambda /\sqrt{2}=r$ and should be compared to the case in the 
Fig.~\ref{fig3} and Fig.~\ref{fig5} with $\lambda =0.225$. The same 
numerical setup for the other parameter apart from $\lambda$ is taken as ones 
in the section 3.1.2. We predict 
\begin{align}
&0.310\lesssim \sin ^2\theta _{12}\lesssim 0.322,\quad 
0.380\lesssim \sin ^2\theta _{23}\lesssim 0.600,\nonumber \\
&0.02\lesssim |J_{CP}|\lesssim 0.04,\quad 
1.5~\text{meV}\lesssim |m_{ee}|\lesssim 2.7~\text{meV},
\end{align}
within $3\sigma $ range of the $\sin\theta _{13}$ from the Fig.~\ref{fig6} (a) to 
\ref{fig6} (d). The $J_{CP}$ is expected to be measured in the future 
long-baseline neutrino experiments. The Fig.~\ref{fig6} (e) is a plot on 
$|r|$--$\lambda $ plane where all points are within $3\sigma $ range of three 
mixing angles. 


\section{Summary}

We have presented neutrino mass matrix textures in the minimal framework of 
the type-I seesaw mechanism where two right-handed Majorana neutrinos are introduced in 
order to reproduce the experimental results of neutrino oscillations. The 
textures can lead to experimentally favored leptonic mixing angles described by 
the TBM and the additional 2-3 rotation. The setup can be 
generically embedded into some scenarios with e.g. the keV sterile neutrino DM.

First, we have presented minimal textures with two zeros in the Dirac neutrino 
mass matrix for the NH case. In the case with two zeros, there are possibly 
three patterns for the position of zero in the Dirac neutrino mass matrix. 
The textures in this case is described by only one free parameter $r$, 
which should be around the ratio between the neutrino mass squared differences. 
We have shown that one of three is marginal to explain the experimental results 
while the others are ruled out because the free parameter is severely restricted. 
The predictions from the possible texture are 
$0.320\lesssim \sin ^2\theta _{12}\lesssim 0.322$, 
$0.660\lesssim \sin ^2\theta _{23}\lesssim 0.680$, 
$0.170~(0.130)\lesssim \sin ^2\theta _{13}~(\sin \theta _{13})\lesssim 
0.196~(0.140)$, and $|J_{CP}|\simeq 0.01$ and $|m_{ee}|\simeq 3.6$ meV within 
$3\sigma $ range of the $\theta _{13}$. 

Next, we have shown (next to minimal) textures with one zero for the NH. 
It has been shown that there are three possible patterns for the position of 
zero as well as the two zero textures but all three textures with one zero lead
 to the same predictions for $\theta _{ij}$, $J_{CP}$, and $m_{ee}$ 
because the same definitions of two model parameters ($r$ and 
$\lambda $) can be taken for all three cases without the loss of generality. 
This is one of different properties between the two and one zero(s) textures. 
We could not predict values of $\theta _{23}$ and $\theta _{13}$ but find a 
clear correlation between $\theta _{12}$ and $\theta _{13}$. The predictions 
from the texture are $0.310\lesssim \sin ^2\theta _{12}\lesssim 0.322$, 
$|J_{CP}|\lesssim 0.04$, and 
$0.8~\text{meV}\lesssim |m_{ee}|\lesssim 3.6~\text{meV}$ within $3\sigma $ range 
of $\theta _{13}$. 

Then, the IH case has been also investigated. In this case, textures with 
zero(s) is conflicts with the experimental results. Therefore, we have 
constructed a texture described by two free parameters without zero. The 
predictions from the texture are $0.310\lesssim \sin ^2\theta _{12}\lesssim 0.322$,
 $0.01\lesssim |J_{CP}|\lesssim 0.04$, and 
$42~\text{meV}\lesssim |m_{ee}|\lesssim 51~\text{meV}$ within $3\sigma $ range of
 the $\theta _{13}$. The result of $|m_{ee}|$ might be tested by the future 
KamLAND-Zen and CUORE experiments of the $0\nu \beta \beta $. We could also find 
some limits for three mixing angles as $0.276\lesssim \sin ^2\theta _{12}$, 
$0.400\lesssim \sin ^2\theta _{23}$, and 
$\sin ^2\theta _{13}~(\sin \theta _{13})\lesssim 0.0784~(0.280)$ in the contrast to 
the NH case with the one zero texture. 

Finally, we have considered the model of the one zero texture with the 
assumptions that one of model parameters is related to the Cabibbo angle or the
 ratio between mass squared differences of the neutrinos. These assumptions are
 now phenomenologically allowed and might be motived for studies of theory 
and/or symmetry in the quark/lepton sectors behind the SM. The first concern 
was the case with the assumption that the model parameter $\lambda $ is the 
Cabibbo angle. In the case, we could not still predict values of the 
$\theta _{23}$ but find $\sin ^2\theta _{12}\simeq 0.315$ and 
$\sin ^2\theta _{13}~(\sin \theta _{13})\simeq 0.0272~(0.165)$. For $J_{CP}$ and 
$|m_{ee}|$, $0.02\lesssim |J_{CP}|\lesssim 0.04$ and 
$1.2~\text{meV}\lesssim |m_{ee}|\lesssim 3.4~\text{meV}$ are predicted within 
$3\sigma $ range of the $\theta _{13}$. In the second one such that the 
parameter is taken as $\lambda/\sqrt{2}=r$, we have found 
$0.310\lesssim \sin ^2\theta _{12}\lesssim 0.322$, 
$0.380\lesssim \sin ^2\theta _{23}\lesssim 0.600$, 
$0.02\lesssim |J_{CP}|\lesssim 0.04$, and 
$1.5~\text{meV}\lesssim |m_{ee}|\lesssim 2.7~\text{meV}$ within $3\sigma $ range 
of the $\theta _{13}$. This case of $\lambda /\sqrt{2}=r$ may be more 
interesting if $r$ and $\lambda $ are just taken as the neutrino mass ratio, 
$r=m_2/m_3=\sqrt{\Delta m_{\text{sol}}^2/\Delta m_{\text{atm}}^2}\simeq 0.16$, 
and as the Cabibbo angle, $\lambda /\sqrt{2}\simeq 0.225/\sqrt{2}\simeq 0.16$, 
respectively.

All our results are expected to be comprehensively tested by combining 
results from more precise determinations of mixing angles, leptonic 
CP-violation searches, and the $0\nu \beta \beta $ experiments in the future.

\vspace{0.5cm}
\noindent
{\bf Acknowledgment}

We thank W.~Rodejohann for useful comments. 
The work of R.T. are supported by Research Fellowships of the Japan Society for
 the Promotion of Science (JSPS) for Young Scientists. M.T. is supported by JSPS
 Grand-in-Aid for Scientific Research, 21340055 and 24654062.

\end{document}